\documentclass[a4paper,11pt]{article}
\usepackage{jheppub} 
\usepackage{lineno}
\usepackage{bm}
\usepackage{slashed}
\usepackage{xpatch}
\usepackage[T1]{fontenc}
\usepackage{braket}
\DeclareMathVersion{ttmath}
\SetSymbolFont{letters}{ttmath}{OT1}{\ttdefault}{m}{n}
\xapptocmd{\tt}{\mathversion{ttmath}}{}{}
\linenumbers
\newcommand{\ar}{{\bf AmpRed~}}
\newcounter{mathcounter}
\newcommand{\mathina}{\addtocounter{mathcounter}{1}{\tt \color{blue} In[\themathcounter]:=}}

\newcommand{\mathinb}{\addtocounter{mathcounter}{1}\texttt{ \color{blue} In[\themathcounter]:=}}
\newcommand{\mathoutb}{\texttt{ \color{blue} Out[\themathcounter]=}}

\nolinenumbers

\title{\boldmath Semi-automatic Calculations of Multi-loop Feynman Amplitudes with AmpRed}

\author{Wen Chen}
\affiliation{Key Laboratory of Atomic and Subatomic Structure and Quantum Control (MOE), Guangdong Basic Research Center of Excellence for Structure and Fundamental Interactions of Matter, Institute of Quantum Matter, South China Normal University, Guangzhou 510006, China}
\affiliation{Guangdong-Hong Kong Joint Laboratory of Quantum Matter, Guangdong Provincial Key Laboratory of Nuclear Science, Southern Nuclear Science Computing Center, South China Normal University, Guangzhou 510006, China}

\emailAdd{chenwenphy@gmail.com}

\abstract{We present a Mathematica package {\bf AmpRed} for the semi-automatic calculations of multi-loop Feynman amplitudes with high efficiency and precision. {\bf AmpRed} implements the methods of integration by parts and differential equations in the Feynman-parameter representation. It allows for the calculations of general parametric integrals (which may not have momentum-space correspondences). Various user-friendly tools for multi-loop calculations, such as those to construct and solve differential equations for Feynman integrals, are provided. It can also deal with tensor algebras in non-relativistic field theories. Interfaces to some packages, like {\tt QGRAF} and FORM, are also provided.
}
\begin{document}
\maketitle
\flushbottom

\section{Introduction}

High-precision calculations are at the forefront of contemporary high-energy physics~\cite{Caola:2022ayt}, of which the perturbative calculation of multi-loop Feynman amplitudes is one of the key components. Modern techniques on multi-loop calculations consist of the reduction of Feynman integrals to the so-called master integrals and the calculation of master integrals.

The first and the most successful technique for multi-loop integral reduction is the integration-by-parts~(IBP) method~\cite{Tkachov:1981wb,Chetyrkin:1981qh}. Feynman integrals of the same family~(integrals with the same set of propagators) are not linearly independent. Instead, they satisfy the IBP identities. All the integrals of the same family can be reduced to a smaller set of integrals (called master integrals) by solving IBP identities according to a prescribed ordering of integrals, a method known as the Laporta algorithm~\cite{Laporta:2000dsw}. It can be shown that the number of master integrals is finite~\cite{Baikov:2005nv,Lee:2008tj,Smirnov:2010hn,Lee:2013hzt}. The Laporta algorithm was implemented in many public packages, including AIR~\cite{Anastasiou:2004vj}, {\tt Reduze}~\cite{Studerus:2009ye,vonManteuffel:2012np}, {\tt FIRE}~\cite{Smirnov:2008iw,Smirnov:2014hma,Smirnov:2019qkx}, and {\tt Kira}~\cite{Maierhofer:2017gsa,Klappert:2020nbg}. Relying on the Gauss elimination for a large sparse linear-equation system, which leads to large intermediate expressions (referred to as intermediate expression swell), the Laporta algorithm becomes extremely expensive on both time and memory for the reduction of complicated multi-loop integrals, especially those with several scales. The performance can be significantly improved by combining with the finite-field method~\cite{Kauers:2008zz,Kant:2013vta,vonManteuffel:2014ixa,Peraro:2016wsq,Klappert:2019emp,Peraro:2019svx}. The efficiency can further be enhanced by refining the IBP systems with techniques like the methods of syzygy equations~\cite{Gluza:2010ws,Schabinger:2011dz,Bohm:2017qme,Boehm:2020zig,Wu:2023upw} and parametric annihilators~\cite{Lee:2014tja,Bitoun:2017nre} and the method of block-triangular form~\cite{Liu:2018dmc, Guan:2019bcx,Guan:2024byi}. Alternatively to the Laporta-like algorithms, the reduction can be achieved by applying symbolic rules obtained through either the Gr\"obner-basis technique~\cite{Baikov:1996iu,Tarasov:1998nx,Gerdt:2004kt,Smirnov:2005ky,Smirnov:2006tz} or solving symbolic IBP identities~\cite{Lee:2012cn,Lee:2013mka}. Some new developments on integral reduction include the method of intersection theory~\cite{Mizera:2017rqa,Mastrolia:2018uzb,Frellesvig:2019uqt}, the method of generating function~\cite{Feng:2022uqp,Feng:2022hyg,Guan:2023avw}, etc.\,.

Many techniques for the calculations of master integrals are available. Such as the Mellin-Barnes method~\cite{Boos:1990rg,Usyukina:1992jd,Smirnov:1999gc,Tausk:1999vh,Czakon:2005rk}, the sector-decomposition method~\cite{Hepp:1966eg,Binoth:2000ps,Borowka:2017idc,Smirnov:2021rhf}, the method of difference equations~\cite{Laporta:2000dsw,Tarasov:1996br,Lee:2009dh,Lee:2015eva}, direct integration~\cite{Brown:2009ta,Panzer:2014caa,vonManteuffel:2014qoa}, and the method of differential equations~\cite{Kotikov:1990kg,Remiddi:1997ny,Gehrmann:1999as}. In the past decade, most state-of-art multi-loop calculations have been based on the differential-equation method. By virtue of the finiteness of the number of master integrals, closed differential-equation systems can be obtained for Feynman integrals, which can be solved either analytically~\cite{Henn:2013pwa,Argeri:2014qva,Henn:2014qga,Lee:2014ioa,Lee:2020zfb} or numerically~\cite{Lee:2017qql,Hidding:2020ytt,Liu:2022chg,Armadillo:2022ugh}. The boundary conditions of the differential equations can be obtained through some iterative algorithms~\cite{Liu:2017jxz,Liu:2022mfb,Hidding:2022ycg,Chen:2023hmk}. Recently, a package based on the differential-equation method for the automatic calculation of Feynman integrals, {\tt AMFlow}~\cite{Liu:2022chg}, which implements the auxiliary-mass-flow method~\cite{Liu:2017jxz,Liu:2020kpc,Liu:2022chg,Liu:2022tji,Liu:2022mfb}, was available.

While the traditional integral reduction and differential equations are carried out in the momentum space, it is found that calculations in the parametric representation are advantageous over the momentum-space methods in several aspects. Tensor integrals can directly be parametrized without doing tensor projection~\cite{Tarasov:1996br}, which allows one to decouple loop integrals from the tensor structures from the very beginning. Contrary to momentum-space integrals, parametric integrals respect Lorentz symmetry at the integrand level. IBP-like linear relations can be found for parametric integrals, which allows for the reduction directly in the parametric representation~\cite{Lee:2014tja,Chen:2019mqc,Chen:2019fzm,Chen:2020wsh}~\footnote{Recently, it was pointed out by ref.~\cite{Artico:2023jrc} that the ideas of IBP and differential equations for parametric integrals were proposed by Regge and collaborators~\cite{Regge:1968rhi,Ponzano:1969tk,Ponzano:1970ch,Regge:1972ns}, pre-dating all the modern multi-loop techniques.}. Parametric IBP identities are simpler than those in the momentum space in the sense of the absence of irreducible scalar products and the non-negativity of the indices. Furthermore, the boundary conditions of differential equations can naturally be determined by matching the asymptotic solutions to the asymptotic expansions~\cite{Beneke:1997zp,Smirnov:1999bza,Smirnov:2002pj,Pak:2010pt,Jantzen:2011nz} of the master integrals. Currently, it is only in the parametric representation that a systematic algorithm for the asymptotic expansions is available~\cite{Pak:2010pt}. Conclusively, it is desirable to implement the parametrization-based methods in practical calculations.

In this paper, we present a Mathematica package \ar for the semi-automatic calculations of Feynman amplitudes. It implements the methods developed in refs.~\cite{Chen:2019mqc,Chen:2019fzm,Chen:2020wsh} to reduce Feynman integrals through the parametric representation, and the method described in ref.~\cite{Chen:2023hmk} to recursively calculate parametric integrals. This paper is organized as follows. In sec.~\ref{sec:Meth}, we review the methods used by \ar. In sec.~\ref{sec:Pack}, the detailed usage of \ar is introduced.

\section{The method}\label{sec:Meth}

\subsection{Parametrization}\label{sec:Par}

We consider scalar Feynman integrals with the following structure
\begin{equation}\label{eq:ScalInt}
J(\lambda_0,~\lambda_1,\dots,\lambda_n)\equiv\int\prod_{i=1}^L\frac{\mathrm{d}^dl_i}{\pi^{d/2}}\frac{w_{\lambda_1}(D_1)w_{\lambda_2}(D_2)\cdots w_{\lambda_m}(D_m)}{D_{m+1}^{\lambda_{m+1}+1}D_{m+2}^{\lambda_{m+2}+1}\cdots D_n^{\lambda_n+1}}~.
\end{equation}

\noindent Here $d=-2\lambda_0$, is the spacetime dimension. Conventionally, we define $d=d_0-2\epsilon$ with $d_0$ the integer dimension. $D_i=\sum_{j,k}A_{i,jk}l_j\cdot l_k+2B_{i,jk}l_j\cdot q_k+C_i$, with $q_i$ some external momenta. The $w$ function is defined by~\cite{Chen:2020wsh}
\begin{equation}
w_\lambda(u)\equiv e^{-\frac{\lambda+1}{2}i\pi}\int_{-\infty}^{\infty}\mathrm{d}x\,\frac{1}{(x-i0^+)^{\lambda+1}}e^{iux}~.
\end{equation}

\noindent We have
\begin{subequations}
\begin{align}
w_0(u)=&2\pi\theta(u)~,\\
w_{-1}(u)=&2\pi\delta(u)~,\\
w_{-2}(u)=&2\pi\delta^\prime(u)~.
\end{align}
\end{subequations}

\noindent The integral $J$ in eq.~(\ref{eq:ScalInt}) can be parametrized by~\cite{Chen:2019mqc,Chen:2019fzm,Chen:2020wsh}
\begin{equation}\label{eq:ParInt}
\begin{split}
J(\lambda_0,\lambda_1,\ldots,\lambda_n)=&s_g^{-\frac{L}{2}}e^{i\pi\lambda_f}\frac{\Gamma(-\lambda_0)}{\prod_{i=m+1}^{n+1}\Gamma(\lambda_i+1)}\int \mathrm{d}\Pi^{(n+1)}\mathcal{F}^{\lambda_0}\prod_{i=1}^{n+1}x_i^{\lambda_i}\\
\equiv&s_g^{-\frac{L}{2}}e^{i\pi\lambda_f}\int \mathrm{d}\Pi^{(n+1)}\mathcal{I}^{(-n-1)}\\
\equiv&s_g^{-\frac{L}{2}}e^{i\pi\lambda_f}I(\lambda_0,\lambda_1,\ldots,\lambda_n)~.
\end{split}
\end{equation}

\noindent Here $\lambda_{n+1}\equiv-(L+1)\lambda_0-1+\sum_{i=1}^m\lambda_i-\sum_{i=m+1}^n(\lambda_i+1)$, $s_g$ is the determinant of the $d$-dimensional spacetime metric, and $\lambda_f=-L\lambda_0-\frac{1}{2}m-\sum_{i=m+1}^n(\lambda_{i}+1)$. The integration measure is $\mathrm{d}\Pi^{(n+1)}\equiv\prod_{i=1}^{n+1}\mathrm{d}x_i\delta(1-\mathcal{E}^{(1)}(x))$, with $\mathcal{E}^{(i)}(x)$ a positive definite homogeneous function of $x$ of degree $i$. The region of integration for $x_i$ is $[0,~\infty)$ if $i>m$ and $(-\infty,~\infty)$ if $i\leqslant m$. $\mathcal{F}$ is a homogeneous polynomial of $x$ of degree $L+1$, defined by $\mathcal{F}=F+Ux_{n+1}$. $U$ and $F$ are Symanzik polynomials, defined by $U(x)\equiv\det{A}$, and $F(x)\equiv U(x)\left(\sum_{i,j=1}^L(A^{-1})_{ij}B_i\cdot B_j-C\right)$, with $A_{ij}\equiv\sum_kx_kA_{k,ij}$, $B_{i}^\mu\equiv\sum_{j,k}x_jq_k^\mu B_{j,ik}$, and $C\equiv\sum_ix_iC_i$. For future convenience, we also define $B_{ij}\equiv\sum_{k}x_kB_{k,ij}$.

Some integrals may be scaleless. For normal loop integrals, scaleless integrals can be identified with the criterion provided in ref.~\cite{Lee:2013mka}. However, this criterion does not always work for phase-space integrals. An example would be the integral $\int\mathrm{d}^dl_1\mathrm{d}^dl_2\,\delta(l_1^2)\delta(l_2^2)\delta((l_1-l_2)^2)\delta((l_1-p)^2)$. For the general parametric integrals in eq.~(\ref{eq:ParInt}), an integral is scaleless if the equation
\begin{equation}
    \sum_{\substack{i\leq m,\\\lambda_i=-1}}\frac{\partial\mathcal{F}}{\partial x_i}\iota_i(x)+\sum_{i=1}^{n+1}\frac{\partial\mathcal{F}}{\partial x_i}x_i\kappa_i(x_{n+1})=0
\end{equation}
\noindent has nontrivial solutions for $\kappa$. Note that $\iota_i$ are functions of $x_j$ with $j\neq i$, while $\kappa_i$ only depend on $x_{n+1}$. This criterion can be proven by combining eq.~(\ref{eq:IBP5}) with the fact that $\hat{z}_i\,I=0$ if $i\leq m$ and $\lambda_i=-1$.

A scalar integral $I(\lambda_0,~\lambda_1,\dots,\lambda_n)$ can be understood as a function of the indices $\lambda$. We may define operators raising or lowering the indices. That is
\begin{subequations}
\begin{align}
\mathcal{R}_iI(\lambda_0,\dots,\lambda_i,\dots,\lambda_n)=&(\lambda_i+1)I(\lambda_0,\dots,\lambda_i+1,\dots,\lambda_n)~,\\
\mathcal{D}_iI(\lambda_0,\dots,\lambda_i,\dots,\lambda_n)=&I(\lambda_0,\dots,\lambda_i-1,\dots,\lambda_n)~,\\
\mathcal{A}_iI(\lambda_0,\dots,\lambda_i,\dots,\lambda_n)=&\lambda_iI(\lambda_0,\dots,\lambda_i,\dots,\lambda_n)~.
\end{align}
\end{subequations}
It is understood that
\begin{equation*}
I(\lambda_0,\dots,\lambda_{i-1},-1,\dots,\lambda_n)\equiv\int \mathrm{d}\Pi^{(n)}\left.\mathcal{I}^{(-n)}\right|_{x_i=0},\quad i=m+1,~m+2,~\cdots,~n.
\end{equation*}
\noindent The product of two operators is defined by the successive applications. That is, $(O_1O_2)I\equiv O_1(O_2I)$. We further define
\begin{align*}
\hat{x}_i=&\left\{
\begin{matrix}
\mathcal{D}_i&,&i=1,~2,\ldots,~m,\\
\mathcal{R}_i&,&i=m+1,~m+2,\ldots,~n+1,
\end{matrix}
\right.\\
\hat{z}_i=&\left\{
\begin{matrix}
-\mathcal{R}_i&,&i=1,~2,\ldots,~m,\\
\mathcal{D}_i&,&i=m+1,~m+2,\ldots,~n+1,
\end{matrix}
\right.\\
\hat{a}_i=&\left\{
\begin{matrix}
-\mathcal{A}_i-1&,&i=1,~2,\ldots,~m,\\
\mathcal{A}_i&,&i=m+1,~m+2,\ldots,~n+1.
\end{matrix}
\right.
\end{align*}
\noindent And we formally define operators $\hat{z}_{n+1}$ and $\hat{x}_{n+1}$, such that $\hat{z}_{n+1}I=I$, and $\hat{x}_{n+1}^iI=\prod_{j=1}^i(\hat{a}_{n+1}+j)I$, with $\hat{a}_{n+1}=-(L+1)\hat{a}_0-\sum_{i=1}^n(\hat{a}_i+1)-1$.~\footnote{In this paper, we use the convention of ref.~\cite{Chen:2023hmk}, which is slightly different from that in refs.~\cite{Chen:2019fzm,Chen:2020wsh}.} We have the following commutation relations:
\begin{subequations}
\begin{align}
\hat{z}_i\hat{x}_j-\hat{x}_j\hat{z}_i=&\delta_{ij},\\
\hat{z}_i\hat{a}_j-\hat{a}_j\hat{z}_i=&\delta_{ij}\hat{z}_i,\\
\hat{x}_i\hat{a}_j-\hat{a}_j\hat{x}_i=&-\delta_{ij}\hat{x}_i.
\end{align}
\end{subequations}

Tensor integrals can be parametrized by using a generator method. A tensor integral is obtained by applying a chain of operators $P_i^\mu$ on a scalar integral. That is,
\begin{equation}\label{eq:ParTens}
\begin{split}
J_{i_1i_2\dots}^{\mu_1\mu_2\dots}(\lambda_0,~\lambda_1,\dots,\lambda_n)\equiv&\int\prod_{i=1}^L\frac{\mathrm{d}^dl_i}{\pi^{d/2}}\frac{w_{\lambda_1}(D_1)w_{\lambda_2}(D_2)\cdots w_{\lambda_m}(D_m)}{D_{m+1}^{\lambda_{m+1}+1}D_{m+2}^{\lambda_{m+2}+1}\cdots D_n^{\lambda_n+1}}l_{i_1}^{\mu_1}l_{i_2}^{\mu_2}\dots\\
=&\left[P_{i_1}^{\mu_1}P_{i_2}^{\mu_2}\cdots J(\lambda_0,~\lambda_1,\dots,\lambda_n)\right]_{p=0}~.
\end{split}
\end{equation}
\noindent The operators $P$ are defined by
\begin{equation}\label{eq:TensGen}
P_i^\mu(p)\equiv-\frac{\partial}{\partial p_{i,\mu}}-\widetilde{B}_i^\mu(\hat{x})+\frac{1}{2}\sum_{j=1}^Lp_j^\mu\widetilde{A}_{ij}(\hat{x}),
\end{equation}
\noindent where $\widetilde{A}_{ij}\equiv\mathcal{D}_0U(A^{-1})_{ij}$ and $\widetilde{B}_i^\mu\equiv\sum_{j=1}^L\widetilde{A}_{ij}B_j^\mu$. The momenta $p_i$ are some auxiliary momenta that are absent in the scalar integral $J$. Because both $\tilde{A}$ and $\tilde{B}$ depend on $\mathcal{D}_0$, they shift the spacetime dimension of the scalar integral $J$.

\subsection{Integral reduction}\label{sec:IntRed}

Like loop integrals in the momentum space, parametric integrals $I$ in eq.~(\ref{eq:ParInt}) satisfy the following identities.
\begin{equation}\label{eq:IBP1}
0=\int \mathrm{d}\Pi^{(n+1)}\frac{\partial}{\partial x_i}\mathcal{I}^{(-n)}+\delta_{\lambda_i0}\theta(i-m-\frac{1}{2})\int \mathrm{d}\Pi^{(n)}\left.\mathcal{I}^{(-n)}\right|_{x_i=0}.
\end{equation}
\noindent By using the operators $\hat{x}$ and $\hat{z}$, we can write these equations in the following form.
\begin{equation}
    \left[\mathcal{D}_0\frac{\partial\mathcal{F}(\hat{x})}{\partial\hat{x}_i}-\hat{z}_i\right]\hat{x}_{n+1}I=0.
\end{equation}
\noindent We assume that $\hat{x}_{n+1}$ is always to the right of $\hat{x}_i$ with $i<n+1$ in $\mathcal{F}(\hat{x})$. This equation can be understood as a polynomial equation of $\hat{x}$. That is
\begin{equation}\label{eq:IBP2}
    \left[\mathcal{D}_0\frac{\partial\mathcal{F}(\hat{x})}{\partial\hat{x}_i}-\hat{z}_i\right]\hat{x}_{n+1}\approx0.
\end{equation}
\noindent Here we use $\approx$ instead of $=$ to indicate that this identity is valid only when applied to nontrivial parametric integrals.

Let $b$ be a positive integer and $f_i(x)\equiv \sum_{a=0}^bf_i^{(a)}x_{n+1}^{b-a}$ be the solutions of the equation
\begin{equation}
    \sum_{i=1}^{n+1}f_i\frac{\partial\mathcal{F}}{\partial x_i}=0~.
\end{equation}
\noindent Then by using eq.~(\ref{eq:IBP2}), we can get
\begin{equation}\label{eq:IBP5}
    \sum_{a=0}^b\left[\left(\sum_{i=1}^{n+1}\hat{z}_if_i^{(a)}(\hat{x})\right)\prod_{j=-b}^{-a}(\hat{a}_{n+1}+1+j)\right]\approx0~.
\end{equation}
\noindent These equations play the roles of IBP identities free of dimensional shift.

The integral reduction in the parametric representation is based on eq.~(\ref{eq:IBP2}). \ar uses two different methods to carry out the integral reduction: method I and method II. In method I, auxiliary external momenta and auxiliary propagators~(and correspondingly, negative indices) are introduced. Integrals with negative indices are defined by~\cite{Lee:2014tja}
\begin{equation}
\begin{split}
&I(\lambda_0,\ldots,\lambda_{i-1},-\Lambda,\lambda_{i+1},\ldots,\lambda_n)
\equiv \lim_{\lambda_i\to-\Lambda}I(\lambda_0,\ldots,\lambda_{i-1},\lambda_i,\lambda_{i+1},\ldots,\lambda_n)\\
=&\frac{(-1)^{\Lambda-1}\Gamma(-\lambda_0)}{\prod_{j=m+1,j\neq i}^{n+1}\Gamma(\lambda_j+1)}\int\mathrm{d}\Pi^{(n)}\,
\partial^{\Lambda-1}_{x_i} \mathcal{F}^{\lambda_0}|_{x_i = 0}
\prod_{j\neq i}^{n+1}x_j^{\lambda_j},\quad \Lambda\in N,~i>m~.
\end{split}
\end{equation}
\noindent This method is closer to the standard momentum-space IBP method. In method II, all the indices of the parametric integrals are non-negative but may be with spacetime dimensions different from that of the original integral in the momentum space.

\subsubsection{Method I}

In method I, we introduce some auxiliary external momenta if the Gram determinant of the external momenta vanishes, and we introduce some auxiliary propagators to make them a complete basis, as in the standard momentum-space IBP method.

We make some definitions first. Let $a_{ij}$ and $b_{ij}$ be the solutions of
\begin{equation}
\sum_jb_{ij}\frac{\partial A(x)}{\partial x_j}=0~,\qquad 
\sum_ja_{ij}\frac{\partial B(x)}{\partial x_j}=0~.
\end{equation}
\noindent In general, solutions of these two equations could be linearly dependent. We denote the linearly dependent part of the solutions by $c_{ij}$. By default, we assume that these solutions are excluded from $a_{ij}$ and $b_{ij}$ and therefore $a_{ij}$ and $b_{ij}$ are linearly independent. For brevity, we denote
\begin{subequations}
\begin{align}
\frac{\partial}{\partial a_i}\equiv\sum_ja_{ij}\frac{\partial}{\partial \hat{x}_j}~,&& \hat{z}_{a_i}\equiv\sum_ja_{ij}\hat{z}_j~,\\
\frac{\partial}{\partial b_i}\equiv\sum_jb_{ij}\frac{\partial}{\partial \hat{x}_j}~,&& \hat{z}_{b_i}\equiv\sum_jb_{ij}\hat{z}_j~,\\
\frac{\partial}{\partial c_i}\equiv\sum_jc_{ij}\frac{\partial}{\partial \hat{x}_j}~,&&
\hat{z}_{c_i}\equiv\sum_jc_{ij}\hat{z}_j~.
\end{align}
\end{subequations} 
\noindent For integrals with a complete set of propagators, there are matrices $\alpha$ and $\beta$ such that
\begin{equation} 
\sum_k\alpha_{ij,k}\frac{\partial A_{mn}}{\partial a_k}=\frac{1}{2}\left(\delta^{im}\delta^{jn}+\delta^{in}\delta^{jm}\right)\,,\qquad 
\sum_k\beta_{im,k}\frac{\partial B_{jn}}{\partial b_k}= \delta^{ij}\delta^{mn}.
\end{equation}
\noindent We further define
\begin{equation} 
\bar{B}_{ij}\equiv \frac{1}{2}\sum_{k,l}g_{jl}\beta_{il,k}\left(\frac{\partial C}{\partial b_k}+\hat{z}_{b_k}\right)~, \qquad 
\bar{A}_{ij}\equiv -\bar{B}_i\cdot\bar{B}_j-\sum_k\alpha_{ij,k}\left(\hat{z}_{a_k}+\frac{\partial C}{\partial a_k}\right)~,
\end{equation} 
\noindent where $g_{ij}$ is the inverse of the Gram matrix $q_i\cdot q_j$. We have
\begin{subequations}
\begin{align}
\left[\bar{A}_{ij},A_{mn}\right]=&-\frac{1}{2}\left(\delta^{im}\delta^{jn}+\delta^{in}\delta^{jm}\right)~,\\
\left[\bar{B}_{im},~B_{jn}\right]=&\frac{1}{2}\delta_{ij}g_{mn}~,\\
\left[\bar{A}_{ij},B_{mn}\right]=&\left[\bar{B}_{ij},A_{mn}\right]=0~.
\end{align}
\end{subequations}

It can be shown that the operator $P$ in eq.~(\ref{eq:TensGen}) can be traded by
\begin{equation}\label{eq:TensGen2}
P_i^\mu\approx-\frac{\partial}{\partial\bar{p}_{i,\mu}}-\bar{B}_i^\mu+\frac{1}{2}\sum_j\widetilde{A}_{ij}\bar{p}_j^\mu~,
\end{equation}
where $\bar{p}_i$ are auxiliary vectors such that $\bar{p}_i\cdot q_j=0$. By combining eqs.~(\ref{eq:ParTens}) and (\ref{eq:TensGen2}), a tensor integral can be expressed in terms of a sum of integrals of the form $f(\bar{B},~\tilde{A})I$, with $I$ a parametric integral and $f$ a polynomial in $\bar{B}$ and $\tilde{A}$. $\bar{B}$ is free of $\mathcal{D}_0$ and commute with $\tilde{A}$. Chains of $\tilde{A}$ can further be reduced by solving the following identities.
\begin{equation}\label{eq:RedAt}
\begin{split}
\widetilde{A}_{i_2j_2}\widetilde{A}_{i_3j_3}\cdots\widetilde{A}_{i_nj_n}\bar{A}_{i_1j_1}\approx&\widetilde{A}_{i_1j_1}\widetilde{A}_{i_2j_2}\cdots\widetilde{A}_{i_nj_n}(\mathcal{A}_0+\frac{E}{2})\\
&-\frac{1}{2}(\widetilde{A}_{i_1i_2}\widetilde{A}_{j_1j_2}+\widetilde{A}_{i_1j_2}\widetilde{A}_{i_2j_1})\widetilde{A}_{i_3j_3}\cdots\widetilde{A}_{i_nj_n}\\
&-\frac{1}{2}(\widetilde{A}_{i_1i_3}\widetilde{A}_{j_1j_3}+\widetilde{A}_{i_1j_3}\widetilde{A}_{i_3j_1})\widetilde{A}_{i_2j_2}\widetilde{A}_{i_4j_4}\cdots\widetilde{A}_{i_nj_n}\\
&-\cdots\\
&-\frac{1}{2}(\widetilde{A}_{i_1i_n}\widetilde{A}_{j_1j_n}+\widetilde{A}_{i_1j_n}\widetilde{A}_{i_nj_1})\widetilde{A}_{i_2j_2}\widetilde{A}_{i_3j_3}\cdots\widetilde{A}_{i_{n-1}j_{n-1}}~,
\end{split}
\end{equation}
\noindent where $E$ is the number of external momenta. The right-hand side of this equation is of degree $n$ in $\widetilde{A}_{ij}$, while the left-hand side is of degree $n-1$. Thus, by solving these identities, we can reduce the degrees of $\widetilde{A}_{ij}$ recursively.

Compared with $\tilde{A}$ and $\tilde{B}$, $\bar{A}$ and $\bar{B}$ are free of $\mathcal{D}_0$. Thus they do not shift the spacetime dimension.

By combining eqs.~(\ref{eq:ParTens}),~(\ref{eq:TensGen2}),~and~(\ref{eq:RedAt}), a tensor integral can be expressed in terms of parametric integrals of the same spacetime dimensions, which can further be reduced by solving the following IBP identities.
\begin{subequations}\label{eq:IBP3}
\begin{align}
\frac{\partial C}{\partial c_i}+\hat{z}_{c_i}\approx&0~,\\
\sum_j\bar{B}_{kj}A_{ik}\approx&B_{ij}~,\\
\sum_k\bar{A}_{ik}A_{kj}\approx&\left(\mathcal{A}_0+\frac{E}{2}\right)\delta_{ij}~.
\end{align}
\end{subequations}
\noindent These IBPs are equivalent to the momentum-space IBPs.

\subsubsection{Method II}

In method II, eq.~(\ref{eq:TensGen}) is used to parametrize tensor integrals. The parametric integrals are expressed in terms integrals of the form $f(\tilde{A},\tilde{B})\,I$ with $f$ a polynomial in $\tilde{A}$ and $\tilde{B}$ and $I$ a parametric integral. $\tilde{B}$ are of degree of $L$ in $\hat{x}$. Chains of $\tilde{B}$ are reduced by using the identity
\begin{equation}\label{eq:IBPB}
\sum_{j,k=1}^L\frac{\partial A_{jk}}{\partial\hat{x}_i}\widetilde{B}_j\cdot\widetilde{B}_k-2\sum_{j=1}^L\frac{\partial B_j}{\partial\hat{x}_i}\cdot\widetilde{B}_j+\mathcal{D}_0\mathcal{A}_0\frac{\partial U}{\partial\hat{x}_i}+\frac{\partial C}{\partial\hat{x}_i}+\hat{z}_i\approx0~.
\end{equation}
\noindent This equation can be understood as a polynomial equation of $\tilde{B}$ (except for the fact that the last term, $\mathcal{D}_i$, does not commute with $\tilde{B}$). Chains of $\tilde{B}$ can be reduced by using the techniques of polynomial reduction~(see sec. 3.2 of ref.~\cite{Chen:2019fzm}). But the reduction is far from complete. The unreduced integrals can further be reduced by applying symbolic rules, which can be obtained from IBP identities.

By virtue of the non-negativity of the indices of parametric integrals, each parametric integral $I(\lambda_0,\lambda_1,\dots)$ can be obtained by applying a product of $\hat{x}$ on a base integral, say, $I(-\frac{d}{2},0,0,\dots)$. Thus, we build a one-to-one correspondence between parametric integrals and monomials of $\hat{x}$. Consequently, a correspondence is built between symbolic IBP identities and polynomial equations of $\hat{x}$. By proper prescription, we can further make the ordering of the monomials consistent with that of the parametric integrals. Thus, instead of working on symbolic IBP identities, we can directly play with polynomial equations of $\hat{x}$, which can be obtained from eq.~(\ref{eq:IBP2}).

Equation (\ref{eq:IBP2}) can be understood as a polynomial equation of $\hat{x}_i$ except for the fact that $\hat{x}_{n+1}$ and $\hat{z}$ do not commute with the rest $\hat{x}_i$. We define
\begin{equation}
    \mathcal{F}(x)\equiv\sum_{i=0}^{L+1}\mathcal{F}^{(L+1-i)}x_{n+1}^{i}~.
\end{equation}
\noindent By shifting $\hat{x}_{n+1}$ to the leftmost, we can write eq.~(\ref{eq:IBP2}) in the following form.
\begin{equation}\label{eq:IBP4}
    -\hat{z}^\prime_i+\sum_{j=0}^{L+1}\hat{z}^\prime_{-j}\mathcal{F}_i^{(L+1-j)}(\hat{x})\approx0~,
\end{equation}
\noindent where
\begin{subequations}
    \begin{align}
        \hat{z}^\prime_{-i}\equiv~&\left[\prod_{j=1}^{i+1}(\hat{a}_{n+1}+j-i-1)\right]\mathcal{D}_0~,\quad i\geq0~,\\
        \hat{z}^\prime_i\equiv~& \hat{a}_{n+1}\hat{z}_i~,\quad 0<i<n+1~,\\
        \hat{z}_{n+1}^\prime\equiv~&\hat{a}_{n+1}+1~,\\
        \mathcal{F}^{(j)}_i(x)\equiv~&\frac{\partial\mathcal{F}^{(j)}(x)}{\partial x_i}~,\quad i\leq n~,\\
        \mathcal{F}_{n+1}^{(i)}(x)\equiv~&(L+2-i)\mathcal{F}^{(i-1)}(x)~.
    \end{align}
\end{subequations}

In principle, one can generate a complete Gr\"obner basis out of eq.~(\ref{eq:IBP4}) for the noncommutative algebra of $\hat{x}$ and $\hat{z}^\prime$. Then the reduction is (almost) a solved problem. However, generating a complete basis for a non-commutative algebra is nearly formidable in practice. Nevertheless, even an incomplete basis can significantly simplify the reduction.

Instead of computing for a non-commutative algebra, we consider the module generated by
\begin{equation}\label{eq:FMod}
    \mathbb{F}_i\equiv\left(\mathcal{F}_i^{(0)}(x),~\mathcal{F}_i^{(1)}(x),\dots,\mathcal{F}_i^{(L+1)}(x),~\overbrace{0,\dots,~0}^{i-1},~-1,~\overbrace{0,\dots,~0}^{n+1-i}\right).
\end{equation}
\noindent It is easy to see that each member of this module corresponds to an element of the ideal generated by eq.~(\ref{eq:IBP4}) (but the reverse is not true). Thus we convert the problem of a non-commutative algebra to a commutative one. However, operations on modules are not supported by some symbolic systems like Mathematica. Thus, \ar uses the following trick~\cite{Chen:2019fzm}. Instead of computing the Gr\"obner basis for the module generated by $\mathbb{F}$ in eq.~(\ref{eq:FMod}), we compute the Gr\"obner basis of the following polynomials:
\begin{subequations}\label{eq:SymbRulGen}
    \begin{align}
        &-z^\prime_i+\sum_{j=0}^{L+1}z^\prime_{-j}\mathcal{F}_i^{(L+1-j)}(x)~, &i=1,~2,\dots,n+1&,\\
        &z^\prime_iz^\prime_j, &i,~j=-L-1,~-L,\dots,n+1&.
    \end{align}
\end{subequations}
\noindent It is easy to convert the obtained basis to polynomial equations in $\hat{x}$. For each equation, by expressing the leading term in terms of the rest terms, we get a symbolic rule, as the ordering of monomials in $\hat{x}$ is consistent with the ordering of the parametric integrals.

Usually, the obtained symbolic rules are not complete for multi-loop integrals. Thus the reduction is incomplete. The unreduced integrals are further reduced by solving IBP identities.

As a summary, in method II, we reduce tensor integrals through the following steps:
\begin{itemize}
    \item[(1)] Parametrize tensor integrals by combining eqs.~(\ref{eq:ParTens})~and~(\ref{eq:TensGen}). Chains of $\tilde{B}$ are partially reduced by using eq.~(\ref{eq:IBPB}).
    \item [(2)] Reduce parametric integrals by using symbolic rules generated by eq.~(\ref{eq:SymbRulGen}).
    \item [(3)] Further reduce the unreduced integrals by solving IBP identities.
\end{itemize}

\subsection{Differential equations}\label{subsec:DiffEquatA}

After the integral reduction, the amplitudes are expressed in terms of master integrals, which are further calculated by using the method of differential equations. In this subsection, we take $m=0$ for integrals in eq.~(\ref{eq:ParInt}). (That is, we do not consider phase-space integrals.)

Let $y$ be a kinematical variable, then it is easy to get the differential equation for a parametric integral:
\begin{equation}\label{eq:Diff2}
    \frac{\partial}{\partial y}I=-\mathcal{D}_0\frac{\partial\mathcal{F}}{\partial y}I~.
\end{equation}
\noindent For an integral of which the propagators are complete, we have
\begin{equation}\label{eq:Diff1}
\begin{split}
    \frac{\partial}{\partial y}\approx&\sum_{ij}\bar{A}_{ij}\frac{\partial A_{ij}}{\partial y}+\sum_{i,j,k,l}q_i\cdot q_j\bar{B}_{ki}\bar{B}_{lj}\frac{\partial A_{kl}}{\partial y}\\
    &-\sum_{i,j,k}\bar{B}_{ij}B_{ik}\frac{\partial q_j\cdot q_k}{\partial y}-2\sum_{i,j,k}q_j\cdot q_k\bar{B}_{ij}\frac{\partial B_{ik}}{\partial y}+\frac{\partial C}{\partial y}~.
\end{split}
\end{equation}

For single-scale integrals, to get a nontrivial differential-equation system, we need to introduce an auxiliary scale. This can be done by inserting a delta function into a parametric integral~\cite{Chen:2023hmk}. That is,
\begin{equation}
    I(\lambda_0,\lambda_1,\ldots,\lambda_n)=\int \mathrm{d}\Pi^{(n+1)}\mathrm{d}y~\delta(y-\mathcal{E}^{(0)}(x))\mathcal{I}^{(-n-1)}~.
\end{equation}
\noindent For simplicity, we choose
\begin{equation}
    \mathcal{E}^{(0)}=\frac{x_i}{x_j}~.
\end{equation}
\noindent Thus we have
\begin{equation}
\begin{split}
    I(\lambda_0,\lambda_1,\ldots,\lambda_n)=&\int \mathrm{d}\Pi^{(n+1)}\mathrm{d}y~\delta(y-\frac{x_i}{x_j})\mathcal{I}^{(-n-1)}\\
    =&\int\mathrm{d}y\int\mathrm{d}\Pi^{(n)}~x_j\left.\mathcal{I}^{(-n-1)}\right|_{x_i=yx_j}\\
    \equiv&\frac{\Gamma(\lambda_i+\lambda_j+2)}{\Gamma(\lambda_i+1)\Gamma(\lambda_j+1)}\int\mathrm{d}y~y^{\lambda_i}I_y~.
\end{split}
\end{equation}

An arbitrary choice of the pair $\{x_i,~x_j\}$ in $\mathcal{E}^{(0)}$ may result in boundary integrals with divergences not regulated by the spacetime dimension. The unregulated divergences can be identified by analyzing the local behavior of the integrand of the parametric integral. We define
\begin{equation}
\mathcal{F}=\sum_{a=1}^{A}\left(C_{\mathcal{F},a}\prod_i^{n+1}x_i^{\Lambda_{ai}}\right)~.
\end{equation}
\noindent A region vector $\bm{k}_r$ is defined by
\begin{subequations}
    \begin{align}
        \sum_{k=1}^{n+1}\Lambda_{ak}k_{r,k}=&0,\quad a\in S_r~,\\
        \sum_{k=1}^{n+1}\Lambda_{ak}k_{r,k}>&0,\quad a\notin S_r~,
    \end{align}
\end{subequations}
\noindent with $S_r$ a subset of $\{1,~2,\cdots,A\}$ such that the cardinal number of $S_r$ is not less than $n+1$. We normalize $\bm{k}_r$ such that
\begin{equation}
    \min_{a\notin S_r}\left\{\sum_ik_{r,i}\Lambda_{ai}\right\}=1~.
\end{equation}
\noindent It can be shown that a parametric integral $I(\lambda_0,\lambda_1,\ldots,\lambda_n)$ is singular (in the sense that there are unregulated divergences) if there is a region $r$ such that~\cite{Heinrich:2021dbf,Chen:2024mbk}
\begin{equation}
    \nu_r\equiv\sum_{i=1}^{n+1}k_{r,i}(\lambda_i+1)\in\mathbb{Z}^-\cup\{0\}~.
\end{equation}

Among all the pairs free of unregulated divergences, we further choose the pair $\{x_i,~x_j\}$ according to the following rules:
\begin{itemize}
    \item [(1)] We choose the pair $\{i,~j\}$ such that the cardinal number of $R_{ij}$ is minimized, with $R_{ij}$ defined by
    \begin{equation}
    R_{ij}=\left\{r|k_{r,i}>k_{r,j}\right\}~.
\end{equation}
    \item[(2)] Among all the pairs satisfying the first rule, we choose the one such that $\max\{N_r|r\in R_{ij}\}$ is minimized, where $N_r$ is the cardinal number of $S_r$.
\end{itemize}

After choosing a specific $\mathcal{E}^{(0)}$, we get the integral $I_y$, which can be calculated by using the differential-equation method. The differential-equation system is solved numerically by using the method described in ref.~\cite{Liu:2022chg}. That is, we substitute $\epsilon$ by pure numbers and restore the $\epsilon$ dependence through fitting. The boundary conditions (chosen at $y=0$) of the differential equations are determined by matching the asymptotic solutions of the differential equations to the asymptotic expansions of the master integrals, which can also be expressed in terms of parametric integrals. Thus the boundary integrals can further be calculated by using the method described in this section. In other words, this method allows us to calculate parametric integrals recursively. The recursive procedure terminates when the $\mathcal{F}$ polynomials of the boundary integrals have at most $n+1$ terms. In this case, the boundary integrals can be evaluated analytically. That is,
\begin{equation}\label{eq:BoundrInt}
\begin{split}
I(\lambda_0,\lambda_1,\ldots,\lambda_n)=&\frac{\Gamma(-\lambda_0)}{\prod_{i=1}^{n+1}\Gamma(\lambda_i+1)}\int \mathrm{d}\Pi^{(n+1)}\mathcal{F}^{\lambda_0}\prod_{i=1}^{n+1}x_i^{\lambda_i}\\
=&\frac{(L+1)\prod_{a=1}^{n+1}\left[\Gamma(\bar{\lambda}_a)C_{\mathcal{F},a}^{-\bar{\lambda}_a}\right]}{\parallel\Lambda\parallel\prod_{i=1}^{n+1}\Gamma(\lambda_i+1)}~,
\end{split}
\end{equation}
with
\begin{equation}
    \bar{\lambda}_a=\sum_{i=1}^{n+1}(\Lambda^{-1})_{ia}(\lambda_i+1)~.
\end{equation}

While the recursive method described above works quite well for integrals with positive defined $\mathcal{F}$ polynomials, it becomes problematic when a $\mathcal{F}$ polynomial has both positive terms and negative terms. In this case, a Feynman parameter may cross a branch point in the region of integration. Generally speaking, it is not easy to determine the branch while a Feynman parameter crosses a branch point. We solve this problem as follows. We replace each negative coefficient of $\mathcal{F}$, denoted by $-C_{\mathcal{F},a}$, by $-yC_{\mathcal{F},a}$, and construct differential equations with respect to $y$. The imaginary part of $y$ should be $i0^+$ due to the $i0^+$ prescription of Feynman propagators. We determine the boundary conditions at $y=0^-$. All the boundary integrals are with positive definite $\mathcal{F}$ polynomials and thus can further be evaluated by using the method described in this subsection.

\subsection{An example}
To illustrate how the methods described in this section are carried out in practical calculations, we demonstrate the calculations of the following sunset integral in detail.
\begin{align*}
J_{1,12}^{\mu_1\mu_2}(-\frac{d}{2},\lambda_1,\lambda_2,\lambda_3)=\pi ^{-D}\int\mathrm{d}^dl_1\mathrm{d}^dl_2\frac{l_1^{\mu_1} l_2^{\mu_2}}{\left(l_1^2\right)^{\lambda_1+1} \left(l_2^2\right)^{\lambda_2+1} \left[\left(l_1+l_2+q\right)^2-m^2\right]^{\lambda_3+1}}~,
\end{align*}
where $q^2=s$. Here the first subscript $1$ in $J_{1,12}$ is just an arbitrary number to distinguish this integral family from others, and the subscripts $12$ correspond to the subscripts of the loop momenta in the numerator.

In the code, this integral can be input by (see sec.~\ref{subsec:FeynmInt})
\begin{align*}
\mathinb~&\texttt{int=TI[\{l[1], l[2]\}, \{{l[1], 0\}, \{l[2], 0\}, \{l[1] + l[2] + p, m}\},}\\
&\qquad\texttt{FV[l[1], a]*FV[l[2], b]];}
\end{align*}
Here we have replaced the indices $\mu_1$ and $\mu_2$ by $a$ and $b$ for simplicity.

The $\mathcal{F}$ polynomial of this integral family reads
\begin{align*}
\mathcal{F}_1~=~m^2 \left(x_1 x_3^2+x_2 x_3^2+x_1 x_2 x_3\right)-s x_1 x_2 x_3+\left(x_1 x_2+x_3 x_2+x_1 x_3\right) x_4~.
\end{align*}
And the matrices $A$, $B$, $\tilde{A}$, and $\tilde{B}$ are
\begin{align*}
A_1=&\begin{pmatrix}
 \hat{x}_ 1+\hat{x}_ 3 & \hat{x}_ 3 \\
 \hat{x}_ 3 & \hat{x}_ 2+\hat{x}_ 3 
\end{pmatrix}~,\\
B_1^\mu=&(\hat{x}_3,\hat{x}_3)~q^\mu~,\\
\tilde{A}_1=&\mathcal{D}_0\begin{pmatrix}
\hat{x}_ 2+\hat{x}_ 3 & -\hat{x}_ 3 \\
 -\hat{x}_3 & \hat{x}_ 1+\hat{x}_ 3
\end{pmatrix}~,\\
\tilde{B}_1^\mu=&\mathcal{D}_0(\hat{x}_2\hat{x}_3,\hat{x}_1\hat{x}_3)~q^\mu~.
\end{align*}

\subsubsection{Reduction with method II}
We consider the reduction with method II first. The integral $J_{1,12}^{\mu_1\mu_2}$ is parametrized by
\begin{align*}
J_{1,12}^{\mu_1\mu_2}=\left[P_1^{\mu_1}P_2^{\mu_2}J_1\right]_{p=0}=\left(\tilde{B}_{1,1}\tilde{B}_{1,2}q^{\mu_1}q^{\mu_2}-\frac{1}{2}\tilde{A}_{1,12}g^{\mu_1\mu_2}\right)J_1~.
\end{align*}
$\tilde{B}_{1,1}\tilde{B}_{1,2}$ can further be reduced to
\begin{align*}
\tilde{B}_{1,1}\tilde{B}_{1,2}\approx\frac{1}{2s}(\hat{z}_1+\hat{z}_2-\hat{z}_3)\left[(1+\hat{a}_0)\mathcal{D}_0U_1+1-\hat{a}_0\right]+\tilde{B}_{1,1}+\tilde{B}_{1,2}+\frac{m^2}{2s}-\frac{1}{2}~.
\end{align*}
We specific to the reduction of the integral $J_{1,12}^{\mu_1\mu_2}(-\frac{d}{2},0,0,0)$. Since all the subsectors of this integral family vanish, we get
\begin{align*}
J_{1,12}^{\mu_1\mu_2}(-\frac{d}{2},0,0,0)=&q^{\mu_1}q^{\mu_2}\left[-\frac{d}{2s}J_1(-\frac{d}{2}-1,0,0,1)+J_1(-\frac{d}{2}-1,1,0,1)+J_1(-\frac{d}{2}-1,0,1,1)\right.\\
&\left.+\frac{1}{2s}(m^2-s)J_1(-\frac{d}{2},0,0,0)\right]+\frac{1}{2}g^{\mu_1\mu_2}J_1(-\frac{d}{2}-1,0,0,1)
\end{align*}

In the code, the above steps are implemented by
\begin{align*}
\mathinb~\texttt{AlphaParametrize[int, GroebnerBasis->GroebnerBasis]}
\end{align*}

The resulting parametric integrals are first reduced by applying symbolic rules generated by eq.~(\ref{eq:SymbRulGen}). In practice, we add a degree bound for efficiency while generating Gr\"obner basis. Integrals that are not reducible by symbolic rules are $J_1(\lambda_0,0,0,0)$, $J_1(\lambda_0,0,0,1)$, and $J_1(\lambda_0,1,0,0)$. The unreduced integrals are further reduced by solving IBP identities generated by eq.~(\ref{eq:IBP2}). The final result is expressed in terms of two master integrals: $J_1(-\frac{d}{2},0,0,0)$ and $J_1(-\frac{d}{2}-1,0,0,0)$.

In the code, the reduction is implemented by
\begin{align*}
\mathinb~\texttt{AlphaReduce[int]}
\end{align*}

\subsubsection{Reduction with method I}
To carry out the reduction with method I, we introduce two auxiliary propagators: $q\cdot l_1$ and $q\cdot l_2$. The $\mathcal{F}$ polynomial of this integral family reads
\begin{align*}
\mathcal{F}_2=&x_1 x_2 x_3 \left(m^2-s\right)+m^2 \left(x_1+x_2\right) x_3^2+\left(x_1 x_2+x_3 x_2+x_1 x_3\right) x_6\\
&~+\frac{s}{4} \left(x_1 x_4^2+x_3 x_4^2+4 x_1 x_3 x_4-2 x_3 x_5 x_4+x_2 x_5^2+x_3 x_5^2+4 x_2 x_3 x_5\right).
\end{align*}
The corresponding matrices $\bar{A}$ and $\bar{B}$ are
\begin{align*}
\bar{A}_2=&\begin{pmatrix}
-\frac{1}{s}\hat{z}_ 5^2-\hat{z}_ 1 & \frac{1}{2} \left(m^2+\hat{z}_ 1+\hat{z}_ 2-\hat{z}_ 3+2 \hat{z}_ 4+2 \hat{z}_ 5\right)-\frac{1}{s}\hat{z}_ 4 \hat{z}_ 5-\frac{1}{2}s \\
 \frac{1}{2} \left(m^2+\hat{z}_ 1+\hat{z}_ 2-\hat{z}_ 3+2 \hat{z}_ 4+2 \hat{z}_ 5\right)-\frac{1}{s}\hat{z}_4 \hat{z}_ 5-\frac{1}{2}s & -\frac{1}{s}\hat{z}_ 4^2-\hat{z}_ 2
\end{pmatrix}~,\\
\bar{B}_2^\mu=&\frac{q^\mu}{s}(\hat{z}_5,\hat{z}_4)~.
\end{align*}
By virtue of eqs.~(\ref{eq:TensGen2} and \ref{eq:RedAt}), we have
\begin{align*}
J_{1,12}^{\mu_1\mu_2}(-\frac{d}{2},0,0,0)=&J_{2,12}^{\mu_1\mu_2}(-\frac{d}{2},0,0,0,-1,-1)\\
=&\left[\bar{B}_{2,1}\bar{B}_{2,2}q^{\mu_1}q^{\mu_2}-\frac{1}{2}\tilde{A}_{2,12}\left(g^{\mu_1\mu_2}-\frac{q^{\mu_1}q^{\mu_2}}{q^2}\right)\right]J_2(-\frac{d}{2},0,0,0,-1,-1)\\
=&\left[\bar{B}_{2,1}\bar{B}_{2,2}q^{\mu_1}q^{\mu_2}-\frac{1}{1-d}\bar{A}_{2,12}\left(g^{\mu_1\mu_2}-\frac{q^{\mu_1}q^{\mu_2}}{q^2}\right)\right]J_2(-\frac{d}{2},0,0,0,-1,-1)~.
\end{align*}
Since $\bar{A}$ and $\bar{B}$ are free of $\mathcal{D}_0$, the resulting parametric integrals are of the same spacetime dimension $d$.

In the code, this step is carried out by
\begin{align*}
\mathinb~\texttt{AlphaParametrize[int, Method->1]}
\end{align*}

The obtained parametric integrals are reduced by solving IBP identities generated by eq.~(\ref{eq:IBP3}), which is implemented in the code by
\begin{align*}
\mathinb~\texttt{AlphaReduce[int, Method->1]}
\end{align*}

\subsubsection{Calculation of master integrals}
The integrals obtained in the previous subsections can be calculated by constructing differential equations with respect to $m$. Nevertheless, to illustrate the method described in sec.~\ref{subsec:DiffEquatA}, we introduce an auxiliary scale $y$ by inserting a delta function $\delta(y-\frac{x_3}{x_4})$. That is~(According to the convention in sec.~\ref{sec:Par}, a parametric integral $I$ differs from the corresponding integral $J$ by only a constant prefactor.)
\begin{align*}
I_1(-\frac{d}{2},0,0,0)=&\frac{1}{\Gamma(\lambda_4+1)}\int\mathrm{d}y\int\mathrm{d}\Pi^{(4)}~\delta(y-\frac{x_3}{x_4})\mathcal{F}_1^{-\frac{d}{2}}x_4^{\lambda_4}\\
=&\frac{1}{\Gamma(\lambda_4+1)}\int\mathrm{d}y\int\mathrm{d}\Pi^{(3)}~\left(\left.\mathcal{F}_1\right|_{x_3\to yx_3,x_4\to x_3}\right)^{-\frac{d}{2}}x_3^{\lambda_4+1}\\
\equiv&(\lambda_4+1)\int\mathrm{d}y~I_3(-\frac{d}{2},0,0)~.
\end{align*}

In the code, this step is implemented by~(see secs.~\ref{subsec:FeynmInt} and \ref{subsec:DiffEquatB})
\begin{align*}
\mathinb~\texttt{int2 = AlphaAddScale[AlphaInt[1, \{-D/2, 0, 0, 0, 0\}], y]}
\end{align*}

The obtained $y$-dependent integral can be calculated by using the differential-equation method. For the integral family $I_3$, there is only one master integral. The differential equation of it reads
\begin{align*}
\frac{\mathrm{d}}{\mathrm{d}y}I_3(-\frac{d}{2},0,0)=&-\frac{1}{2 (3 d-2) y \left(m^2 y+1\right) \left(m^2 y-s y+1\right)}I_3(-\frac{d}{2},0,0)\\
&\times\left[\left(9 d^2-16 d+4\right) m^2 y^2 \left(m^2-s\right)+2 (d-2) (3 d-2)\right.\\
&\left.+y \left(\left(15 d^2-32 d+12\right) m^2+\left(4-3 d^2\right) s\right)\right]~.
\end{align*}

In the code, the differential-equation system is obtained by
\begin{align*}
\mathinb~\texttt{des = AlphaDES[AlphaInt[3, \{-D/2, 0, 0, 0\}], y]}
\end{align*}

The boundary conditions of the differential-equation system can be determined by matching the asymptotic solution of the differential equation to the asymptotic expansion of the master integral, which are
\begin{align*}
I_{3,1}=&Cy^{2\epsilon-2}~,\\
I_{3,2}=&y^{2 \epsilon -2}I_4(-\frac{d}{2},0,0)~.
\end{align*}
The $\mathcal{F}$ polynomial of the integral family $I_4$ is
\begin{equation*}
\mathcal{F}_4~=~x_1 x_3^2+x_2 x_3^2+x_1 x_2 x_3~,
\end{equation*}
which has exactly $3$ terms. Thus it can be calculated by using eq.~(\ref{eq:BoundrInt}). That is
\begin{align*}
I_4(-\frac{d}{2},0,0)~=~\frac{\Gamma (1-\epsilon )^2 \Gamma (\epsilon )}{\Gamma (4-3 \epsilon )}~.
\end{align*}

In the code, the asymptotic solution and the asymptotic expansion are obtained by
\begin{align*}
\mathinb~&\texttt{asy1 = DESAsymptoticSolve[des[[2]], \{y, 0, 0\}];}\\
&\texttt{asy2 = AlphaSeries[des[[1, 1]], y -> 0]//AlphaIntEvaluate;}
\end{align*}
Here {\tt AlphaSeries} carries out the asymptotic expansion and {\tt AlphaIntEvaluate} evaluates the obtained integral in terms of gamma functions.

After fixing the boundary conditions, the differential-equation system can be solved either analytically or numerically.

\section{The package}\label{sec:Pack}

\ar can be downloaded from

\href{https://gitlab.com/chenwenphy/ampred.git}{https://gitlab.com/chenwenphy/ampred.git}

\noindent It can be run under a Wolfram Mathematica of version 10 or newer. No installation is needed. However, the dependencies must be correctly installed if one needs to use the interfaces~(see sec.~\ref{subsec:Int}). And the interfaces only work under Linux-like systems. As far as \ar is located in a default directory of Mathematica, it can be loaded by running\\

\mathinb {\tt ~<<AmpRed`}\\

\noindent Various examples on using \ar are provided in the folder {\tt AmpRed/examples}~(see sec.~\ref{subsec:Exampl}).

All the global options of \ar are controlled by the function {\tt AmpRed}. {\tt AmpRed[\,]} gives the version, path, and a list of names of \ar. {\tt AmpRed} has the following options:
\begin{itemize}
    \item {\tt "LoadNR"} specifies whether to load {\tt NRalgebra} or not~(see sec.~\ref{subsec:AlgNonrFieldTheor}). Similarly, {\tt "LoadAux"} specifies whether to load the auxiliary file or not~(see sec.~\ref{subsec:AuxFunct}). By default, {\tt "LoadNR"->\\
    False} and {\tt "LoadAux"->True}.
    \item {\tt MemoryConstraint} specifies the maximum memory that is allowed to be used by some functions, such as {\tt AlphaReduce}. While reaching the memory constraint, the computation will be aborted.
    \item {\tt TimeConstraint} specifies the maximum time that is allowed to be spent on some operations by some functions, such as generating the symbolic rules while running {\tt AlphaReduce}.
    \item {\tt "SpaceTimeDimension"} specifies the space-time dimension, which can be used by some functions, such as {\tt ContractLI} and {\tt FeynmanInt}. By default, {\tt "SpaceTimeDimension"\\
    ->D}.
    \item {\tt "MetricSignature"} specifies the metric signature. By default, {\tt "MetricSignature"\\
    ->\{1,3\}}. Note that for 3-dimensional Euclidean space, one needs to set \\
    {\tt "MetricSignature"->\{0,3\}} (rather than {\tt "MetricSignature"->\{3,0\}}).
    \item  {\tt "LoopPrefactor"} specifies the prefactor for each fold of loop integration. The value will be passed to the variable {\tt \$LoopPrefactor}. By default, {\tt "LoopPrefactor"->\\
    -I/Gamma[1+eps]}.
    \item {\tt "SpaceDimension"} specifies the default space dimension. By default, {\tt "SpaceDimension"\\
    ->D-1}. This option works only when {\tt "LoadNR"->True}.
    \item {\tt Directory} gives the current working directory. If the variable {\tt \$UseDisk = True}, the data will be saved on the disk in this directory. By default {\tt \$UseDisk = True}, and the option value of {\tt Directory} is the directory of the notebook.
    \item {\tt FileBaseName} gives the file name of the notebook. If the Wolfram System is not being used with a notebook-based front end, a temporary file name will be assigned. If {\tt \$UseDisk = True}, all the data will be saved in a folder with this name.
    \item {\tt Path} specifies a list of directories where executables can be found by \ar. For example, if one needs to use the FORM interface~(see sec.~\ref{subsec:Int}), and FORM is not installed in a default directory, one can add the directory of the FORM executable to the option value of {\tt Path}.
\end{itemize}
\noindent If one resets options for {\tt AmpRed}, it is suggested to rerun {\tt AmpRed[]}. Note that some options like {\tt "LoadNR"} should be set before loading \ar. This can be done by setting the variable {\tt \$AROptions} before loading \ar. For example,
\begin{align*}
    \mathinb&\texttt{\$AROptions=\{"LoadNR"->True, "LoadAux"->False\};}\\
    \addtocounter{mathcounter}{1}
    &\texttt{<<AmpRed`}
\end{align*}

The usage of some public functions will be introduced in this section.

There are some variables and functions frequently used within \ar:

\begin{itemize}
    \item {\tt \$UseDisk} specifies whether to save the data on the disk to save memory or not. By default, {\tt \$UseDisk=True}.
    \item {\tt \$LoopPrefactor} is the prefactor of each fold of loop integration.
    \item {\tt ApartG} carries out the partial fraction by using an algorithm similar to that described in ref.~\cite{Heller:2021qkz}. The usage of {\tt ApartG} is the same as that of the built-in function {\tt Apart}~(except for the options). Note that {\tt ApartG} may be very time-and-memory-consuming for large expressions. Thus, it is suggested to divide a large expression into some subexpressions and compute them one by one.\\
    {\tt ApartG} has the following options:\\
    {\tt Factor} specifies the function to factor polynomials. By default {\tt Factor->Factor}.\\
    {\tt FactorList} is a list of factors that can be used to factor polynomials.\\
    {\tt GroebnerBasis} specifies the function to generate Gr\"obner basis. If a user-defined function is used, one needs to ensure that it is of the same usage as that of the built-in function {\tt GroebnerBasis}.\\
    {\tt PolynomialReduce} specifies the function to reduce polynomials.\\
    {\tt Polynomials} is a list of polynomials that vanish.\\
    {\tt Parallelize} specifies whether to parallelize the computation or not.
    \item {\tt ExpandAR[exp, x]} expands {\tt exp} and wrap terms not free of {\tt x} with a head {\tt HoldAR}. {\tt x} could be a pattern or a list of patterns. If {\tt x} is omitted, every term will be wrapped. {\tt HoldAR[1]} will be replaced by {\tt 1} if one sets the option {\tt OneIdentity->True}.
\end{itemize}

\subsection{Tensor algebras}\label{sec:TensAlg}

\begin{itemize}
    \item \ar uses a convention similar to that of FeynCalc~\cite{Mertig:1990an,Shtabovenko:2016sxi}. That is, it uses {\tt Momentum[\_]} and {\tt LorentzIndex[\_]} to represent a momentum and a Lorentz index. Note that {\tt Plus} is not allowed inside {\tt Momentum}. The sum of two momenta $p$ and $q$ can be represented by {\tt Momentum[p]+Momentum[q]}.\\
    {\tt Pair[LorentzIndex[a],Momentum[p]]} represents a vector $p^a$.\\
    {\tt Pair[Momentum[p],Momentum[q]]} represents the inner product $p\cdot q$.\\
    {\tt Pair[LorentzIndex[a],LorentzIndex[b]]} represents the metric $g^{ab}$.\\
    {\tt LeviCivita[LorentzIndex[a],LorentzIndex[b],LorentzIndex[c],LorentzIndex[d]]} represents a Levi-Civita tensor $\epsilon^{abcd}$.\\
    Some auxiliary functions are provided, which allow us to input a vector, an inner product, and a metric by {\tt FV[p,a]}, {\tt~SP[p,q]}, and {\tt MT[a,b]}, respectively.\\
    {\tt LorentzIndex}, {\tt Momentum}, and {\tt Pair} are not protected. Thus users can freely set the downvalues of them~(but not ownvalues).
    \item A chain of spinors and gamma matrices is represented by {\tt SpinorChain[\_\_]}.\\
    {\tt DiracSpinor[p,1,I]} represents a Dirac spinor $u(p)$ for a fermion with momentum p. {\tt DiracSpinor[p,-1,I]} represent a Dirac spinor $v(p)$ for an anti-fermion. {\tt DiracSpinor[p,1,-I]} and {\tt DiracSpinor[p,-1,-I]} represent $\bar{u}(p)$ and $\bar{v}(p)$ respectively.\\
    {\tt DiracGamma[LorentzIndex[a]]} and {\tt DiracGamma[Momentum[p]]} represent Dirac gamma matrices $\gamma^a$ and $\slashed{p}$ respectively.\\
    The trace of a chain of Dirac matrices is represented by {\tt SpinorChain[Bra,\_\_,Ket]}.\\
    One can input $\gamma^a$ and $\slashed{p}$ by {\tt GA[a]} and {\tt GS[p]} for simplicity.
    Here are some examples:
    \begin{align*}
    \mathinb&~\texttt{SpinorChain[GS[p],DiracSpinor[p,1,I]]}\\
    \mathoutb&~\bar{p}.u(p)\\
    \mathinb&~\texttt{SpinorChain[Bra,GA[a],GS[p],Ket]}\\
    \mathoutb&~\braket{\gamma^a.\bar{p}}
    \end{align*}
    \item {\tt MomentumExpand} expands {\tt Plus} in {\tt Pair} and {\tt DiracGamma}.
    \item {\tt PolarizationVector[p,I]} represents a polarization vector with momentum p. \\
    {\tt PolarizationVector[p,-I]} represents the Hermite conjugation of a polarization vector.
    \item {\tt ContractLI[exp]} contracts paired Lorentz indices in {\tt exp}.
    \item {\tt SpinorChainSimplify[exp]} simplifies chains of Dirac spinors and Dirac gamma matrixes in {\tt exp}, including calculating the traces of gamma matrices.\\
    By default, {\tt SpinorChainSimplify} uses the anti-commutative $\gamma^5$ scheme. One can fix the ordering of a trace of gamma matrices by inserting a {\tt Hold[1]} into the spinor chain. {\tt SpinorChainSimplify} will automatically shift {\tt Hold[1]} to the leftmost of a trace.
    \item {\tt HermiteConjugate[exp]} gives the Hermite conjugation of {\tt exp}.
    \item {\tt PolarizationSum[exp]} sums {\tt exp} over the polarizations of spinors and polarization tensors (vectors) in {\tt exp}.
    \item {\tt SquareAmplitude[A]} calculates the squared amplitude $|A|^2$.\\
    \item {\tt ColourIndex[r,i]} with {\tt r} a string represents a colour index {\tt i} of the representation {\tt r}. \ar uses {\tt "F"} to represent the fundamental representation and {\tt "A"} to represent the adjoint representation. A colour matrix $T_{r,ij}^a$ of the representation {\tt r} is represented by {\tt ColourChain[ColourIndex[r,i],ColourIndex["A",a],ColourIndex[r,j]]}. A chain of colour matrices, like $T^a_{r,ik}T^b_{r,kj}$, is represented by {\tt ColourChain[ColourIndex[r,i],}\\
    {\tt ColourIndex["A",a],ColourIndex["A",b],ColourIndex[r,j]]}. Note that \ar expresses the structure constant of a SU(N) group in terms of a colour matrix of the adjoint representation.\\
    \item {\tt ColourN[r]} represents the dimension of the colour representation {\tt r}. And {\tt \$ColourN} represents the dimension of the fundamental representation.
    \item {\tt ColourCasimir[1, r]} represents the Casimir operator in the representation {\tt r}.\\
    {\tt ColourCasimir[2, r]} represents the quadratic Casimir operator.
    \item {\tt ColourSimplify[exp]} simplifies chains of colour matrices in {\tt exp}.
\end{itemize}

\subsection{Feynman integrals}\label{subsec:FeynmInt}
\begin{itemize}
    \item {\tt eps} represents $\epsilon\equiv\frac{1}{2}(d_0-d)$, with $d$ the spacetime dimension and $d_0$ the integer spacetime dimension.
    \item {\tt FeynmanInt[\{l$_1$,l$_2$\},~\{\{D$_1$,i$_1$\},\{D$_2$,i$_2$\},\dots\},~\{\{W$_1$,j$_1$\},\{W$_2$,j$_2$\},\dots\}]} represents a Feynman integral $\int\frac{\mathrm{d}^dl_1}{\pi^{d/2}}\frac{\mathrm{d}^dl_2}{\pi^{d/2}}\cdots\,D_1^{i_1}D_2^{i_2}\cdots w_{j_1}(W_1)w_{j_2}(W_2)\cdots$.\\
    {\tt TensorInt[\{l$_1$,l$_2$,\dots\},~\{\{p$_1$,m$_1$,i$_1$\},\{p$_2$,m$_2$,i$_2$\},\dots\},~\{\{W$_1$,j$_1$\},\{W$_2$,j$_2$\},\dots\},~Num]} represents a tensor integral $\int\frac{\mathrm{d}^dl_1}{\pi^{d/2}}\frac{\mathrm{d}^dl_2}{\pi^{d/2}}\cdots\,\frac{w_{j_1}(W_1)w_{j_2}(W_2)\cdots}{(p_1^2-m_1^2)^{i_1}(p_2^2-m_2^2)^{i_2}\cdots}\text{Num}$, with {\tt Num} the numerator.\\
    {\tt ToTensorInt[exp]} converts {\tt FeynmanInt} in {\tt exp} to {\tt TensorInt}.\\
    {\tt TensorIntExplicit[exp]} converts {\tt TensorInt} in {\tt exp} to {\tt FeynmanInt}.
    \item {\tt ToFeynmanInt[exp,~\{l$_1$,l$_2$,\dots\}]} expresses Feynman integrals in {\tt exp} in terms of {\tt FeynmanInt}. The integration measure is added through the option {\tt "Measure"}. For example,\\
    \begin{align*}
        \mathinb&\texttt{ToFeynmanInt[1/(SP[l]*SP[l+p]),\{l\},"Measure"->\{\{SP[l]-m\^~2,-1\}\}]}\\
        \mathoutb&-\frac{i\pi^{-D/2}}{\Gamma(\epsilon+1)}\int d^Dl\frac{2\pi\delta(l^2-m^2)}{l^2(l+p)^2}
    \end{align*}
    Similar to {\tt ApartG}, it is suggested to divide large expressions into subexpressions while running {\tt ToFeynmanInt}.
    \item {\tt AlphaInt}[{\tt n},~\{$\lambda_0$,$\lambda_1$,\dots,$\lambda_{n+1}$\}] represents a parametric integral of the {\tt n}th family with indices $\{\lambda_0,\lambda_1,\dots,\lambda_{n+1}\}$, where $\lambda_i$ are those appeared in the parametric integral in eq.~(\ref{eq:ParInt}). In eq.~(\ref{eq:ParInt}), the index $\lambda_{n+1}$ is not explicitly specified because it is not independent of the other $\lambda_i$. In the code, one may input $\lambda_{n+1}$ with an arbitrary number (it can be automatically corrected). \ar label each integral family with an integer {\tt n}. Information related to an integral family {\tt n} can be obtained with the function {\tt AlphaInfo} (see below).
    \item {\tt AlphaInfo[n, k]} with {\tt n} an integer and {\tt k} a string gives the information related to integral family {\tt n}. For examples, {\tt AlphaInfo[n,"F"]} give the $\mathcal{F}$ polynomial, and {\tt AlphaInfo[n,"P"]} gives the propgators.
    \item {\tt NewAlphaInt[\{F,z[i$_1$],z[i$_2$],\dots,x[i$_{m+1}$],x[i$_{m+2}$],\dots,x[i$_{n+1}$]\},~\{n$_0$,n$_1$,\dots\}]} generate a parametric integral with Feynman parameters $\{z[i_1],z[i_2],\dots,$\\
    $x[i_{m+1}],x[i_{m+2}],\dots,x[i_{n+1}]\}$ and indices $\{n_0,n_1,\dots\}$. {\tt F} is the $\mathcal{F}$ polynomial.
    \item {\tt AlphaIntExplicit[exp]} converts all the {\tt AlphaInt} in exp to their explicit forms.
    \item {\tt AlphaParametrize[exp]} parametrize all the Feynman integrals in {\tt exp}.\\
    {\tt AlphaParamtrize} has the following options:\\
    {\tt Method} specifies the method to parametrize Feynman integrals. The option value could be either {\tt 1} or {\tt 2}, corresponding to method I and II respectively.\\
    {\tt "Replacements"} is a list of replacement rules that will be applied to the $\mathcal{F}$ polynomials.\\
    {\tt "UseForm"} specifies whether to use FORM or not.\\
    {\tt "Regulators"} is a list of regulators for the indices. By default, {\tt "Regulators"->\\
    Automatic}, indicating that all the symbols appearing in the indices are treated as regulators.\\
    {\tt "SymbolicIndices"} specifies symbols appearing in the indices that are not regulators.\\
    {\tt Momentum} specifies a list of external momenta, which is needed when the Gram determinant vanishes and Method I is used.\\
    {\tt Simplify} specifies the function to simplify rational functions.\\
    {\tt GroebnerBasis} specifies the function to generate Gr\"obner basis while generating symbolic rules. It can be a user-defined function, such as {\tt SingularGroebnerbasis} defined in the auxiliary file. By default {\tt GroebnerBasis->None}, indicating that symbolic rules are not generated.\\
    {\tt AlphaSave} specifies whether the cache will be saved in the disk or not. The option value can be {\tt True}, {\tt False}, or a string. A string specifies the directory to save.
    \item {\tt AlphaToFeynman[exp]} converts {\tt AlphaInt} in {\tt exp} to {\tt FeynmanInt} when it is possible. {\tt AlphaToFeynman} does the conversion with the information provided by {\tt AlphaParametrize}.
    \item {\tt AlphaReduce[exp]} reduces all the Feynman integrals in {\tt exp} to master integrals by using the methods described in sec.~\ref{sec:Meth}.\\
    Besides those inherited from {\tt AlphaParametrize}, {\tt AlphaReduce} has the following options:\\
    {\tt Bounds} should be a list of non-negative integers, which specifies the bounds of the nonnegative indices and the negative indices respectively while generating explicit IBP identities.\\
    {\tt PolynomialReduce} specifies the function to carry out the polynomial reduction.\\
    {\tt Rule} specifies whether symbolic rules will be applied or not. The option value should be either {\tt True} or {\tt False}.\\
    \item {\tt AlphaIntEvaluateN[exp, n, rul]} evaluates Feynman integrals in {\tt exp} numerically up to order {\tt n} in {\tt eps}. {\tt rul} is a list of rules for the numerical calculation, which can be omitted if there's no symbol. {\tt AlphaIntEvaluateN[exp, n, rul]} uses {\tt AlphaDES} to construct differential equations and {\tt DESSolveN} to solve the differential-equation system numerically~(see sec.~\ref{subsec:DiffEquatB}).\\
    Besides those inherited from {\tt AlphaDES} and {\tt DESSolveN}, {\tt AlphaIntEvaluateN} has the following options:\\
    {\tt "AnalyticContinuation"} specifies whether to use the method described by the end of sec.~\ref{subsec:DiffEquatA} to do the analytic continuation for complex integrals or not.\\
    {\tt "NumericalReduction"} specifies whether to carry out the reduction numerically or not if {\tt rul} is nonempty. If one needs to change the values of some variables, it is suggested to set {\tt "NumericalReduction"->False}, but then the reduction may be much slower.\\
    {\tt Variables} specifies the head of new variables introduced while generating differential equations. By default, {\tt Variables->y}. If the symbol {\tt y} is already used, it is suggested to use another variable.
    \item {\tt AlphaRegions[int, \{\{y$_1$, n$_1$\}, \{y$_2$, n$_2$\},\dots\}]} returns all the asymptotic regions of the integral {\tt int}. {\tt n$_i$} is the scaling of the parameter {\tt y$_i$}. {\tt AlphaRegions} uses {\tt Qhull}~\cite{Barber:1996} to compute convex hulls.\\
    \item {\tt AlphaSeries[I, \{x, x$_0$, n\}, \{\{y$_1$,n$_1$\}, \{y$_2$,n$_2$\},\dots\}]} expands an {\tt AlphaInt} {\tt I} in power series of {\tt x} around {\tt x=x$_0$} to order {\tt n}. {\tt n$_i$} is the scaling of the kinematic variable {\tt y$_i$}. That is, $y_i\sim (x-x_0)^{n_i}$. The last argument can be omitted if there are no variables other than {\tt x}.\\
    {\tt AlphaSeries[I, x->x$_0$, \{\{y$_1$,n$_1$\}, \{y$_2$,n$_2$\},\dots\}]} only keeps the leading term for each region.
    \item {\tt AlphaSingularQ} checks whether a parametric integral is singular (due to unregulated divergences) or not.
    \item {\tt AlphaSave[dir]} with {\tt dir} a string saves the cache in the directory {\tt dir}.\\
    {\tt AlphaSave[]} saves the cache in the default directory.
    \item {\tt AlphaLoad[dir]} loads the cache saved in the directory {\tt dir}.
    \item {\tt AlphaClear[]} clears all the cache in RAM. {\tt AlphaClear[n]} clears all the cache related to the {\tt n}th integral family.
\end{itemize}

\subsection{Differential equations}\label{subsec:DiffEquatB}

\begin{itemize}
    \item {\tt AlphaD[exp, x]} gives the differentiation of {\tt exp} with respect to {\tt x}. The option {\tt Method} specifies the method to carry out the differentiation. The option value should be either {\tt 1} or {\tt 2}, corresponding to eq.~(\ref{eq:Diff1}) and eq.~(\ref{eq:Diff2}) respectively. By default, {\tt Method->2}.
    \item {\tt AlphaDES[I$_0$, x]} constructs differential equations for a list of Feynman integrals $I_0$. It returns $\{I, M\}$, with $I$ a list of master integrals and $M$ a matrix such that $\frac{\partial I}{\partial x}I=M\cdot I$. {\tt AlphaDES} uses {\tt AlphaD} to carry out the differentiation.\\
    In addition to those that inherit from {\tt AlphaD}, {\tt AlphaDES} has the following options:\\
    {\tt AlphaBasis} specifies whether to use {\tt AlphaBasis} to choose the basis or not.\\
    {\tt AlphaReduce} specifies the function to carry out the integral reduction. It can be a user-defined function, such as {\tt DoKira}~(see sec.~\ref{subsec:Int}).
    \item {\tt AlphaAddScale[I, y, \{i,j\}]} adds an auxiliary scale {\tt y} to the parametric integral {\tt I} through the replacement $x_i\to yx_j$. If the last argument is omitted, {\tt AlphaAddScale} chooses the pair according to the algorithm described in sec.~\ref{subsec:DiffEquatA}.
    \item {\tt DESBoundary[M, x, x$_0$]} gives the boundary conditions that need to be fixed for the differential-equation system $\frac{\partial I}{\partial x}=M\cdot I$. $x_0$ can be omitted if $x_0=0$. The output is a list of the form $\{\{i, n\},\dots\}$, with $\{i, n\}$ indicating that the series coefficient of $(x-x_0)^n$ for the ith integral needs to be calculated.\\
    {\tt DESBoundary[\{I, M\}, x, x$_0$]} gives the boundary conditions that need to be fixed for the differential equation system $\frac{\partial I}{\partial x}=M\cdot I$, with $I$ a list of parametric integrals.
    \item {\tt DESAsymptoticSolve[M, \{x, x$_0$, n\}]} gives the asymptotic solution of the differential-equation system $\frac{\partial I}{\partial x}=M\cdot I$ to order $(x-x_0)^n$.
    \item {\tt DESSolveN[M, \{x, x$_1$, x$_2$\}, n, bnd]} solves the differential-equation system $\frac{\partial f}{\partial x}=M\cdot f$ numerically, and returns the value of $f$ at {\tt x=x$_2$}. The result is expanded to order {\tt n} in {\tt eps}. {\tt bnd} is a list of boundary conditions at {\tt x=x1}. The elements of {\tt bnd} should be of the form {\tt \{i, o\}->c}, which indicates that the series coefficient of $f_i$ of the order {\tt $(x-x1)^o$} term is {\tt c}. {\tt c} should be free of poles in {\tt eps}.\\
    The boundary conditions can be determined through bootstrap by setting the option {\tt "Constraints"}, of which the option value is a list of rules of the form {\tt \{i,x$_0$\}->y$_0$}, indicating that $f_i(x_0)=y_0$.\\
    {\tt DESSolveN} has the following options:\\
    {\tt "Braches"} specifies the branches while crossing branch points. The option value should be a list of rules of the form {\tt x$_0$ -> s}, which indicates that a monomial $(x-x_0)^i$ with $x<x_0$ should be understood as $(x_0-x)^ie^{is\pi}$. By default, $s=-1$.\\
    {\tt "OmittedSingularities"} specifies a list of fake singularities to be omitted.\\
    {\tt PrecisionGoal} specifies the precision goal. If the obtained result does not reach the expected precision, try manually setting the option values of {\tt WorkingPrecision} and {\tt eps}.\\
    {\tt "IntegrationPath"} is a list of points between {\tt x$_1$} and {\tt x$_2$}, around which the differential equations are solved as series expansions. The order of a series expansion is estimated according to the precision goal.\\
    {\tt "ReturnBoundaryConditions"} specifies whether to return the boundary conditions or not.
    \item {\tt GPolyLog[\{\{i\},\{j\},...\},x]} represents a multiple polylogarithm~\cite{Goncharov:1998kja} $G(i,j,\dots,x)$.\\
    {\tt GPolyLog[\{...,\{i,j,...\},...\},x]} represents $G(\dots,i,\dots,x)+G(\dots,j,\dots,x)+\cdots$.
    \item {\tt CIteratedInt[\{a$_1$,a$_2$,\dots\},x]} represents a Chen iterated integral~\cite{Chen:1977oja} \\
    $\int\mathrm{d}\log a_1\int\mathrm{d}\log a_2\cdots$. The integration path is specified through the option {\tt Path}. For example, a straight line from $x=0,~y=0$ to $x=1,~y=2$ is represented by {\tt Path->\{\{x->0,y->0\},\{x->1,y->x\}\}}.
    \item {\tt CIIToGPL[exp]} converts {\tt CIteratedInt} in {\tt exp} to {\tt GPolyLog}.
    \item {\tt DysonSolve[M, \{x, x0, ord\}]} with {\tt M} a square matrix gives the solution of the differential equation $\frac{\partial U(x_0,x)}{\partial x}=M\cdot U(x_0,x)$ in terms of a Dyson series, with $U(x_0,x)$ a squared matrix such that $U(x_0,x_0)=\mathbb{I}$. Elements of {\tt M} should be either a rational function of {\tt x} or linear combinations of terms of the form {\tt DifferentialD[Log[f]]} with {\tt f} functions of {\tt x}.
    \item {\tt GPLIntegrate[exp,\{x,x$_1$,x$_2$\}]} integrate {\tt exp} with respect to {\tt x} according to the definition of multiple polylogarithms.
    \item {\tt GPLSeries[exp,~\{x,x$_0$,n\}]} expands {\tt GPolyLog[\dots]} in {\tt exp} in power series to the order $(x-x_0)^n$.
    \item {\tt GPolyLogN[exp]} evaluates {\tt GPolyLog} and {\tt CIteratedInt} in {\tt exp} numerically.
\end{itemize}

\subsection{Algebras in non-relativistic field theories}\label{subsec:AlgNonrFieldTheor}

Some functions for algebras in non-relativistic field theories are defined in the file\\
{\tt "AmpRed/nralgebra.m"}. To use these functions, one needs to append {\tt "LoadNR"->True} to {\tt \$AROptions} before loading \ar. These functions are:

\begin{itemize}
    \item {\tt IndexNR} is the non-relativistic version of {\tt LorentzIndex}.
    \item {\tt MomentumNR} is the non-relativistic version of {\tt Momentum}.
    \item {\tt PairNR} is the non-relativistic version of {\tt Pair}.
    \item {\tt LeviCivitaNR[a,b,c]} is the three-dimensional Levi-Civita tensor $\epsilon^{abc}$.
    \item {\tt WeylSpinor} represents a Weyl spinor. It is of the same usage as that of {\tt DiracSpinor}.
    \item {\tt PauliSigma[IndexNR[i]]} represents a Pauli sigma matrix $\sigma^i$, and {\tt PauliSigma[MomentumNR[p]]} represents $\bm{p}\cdot\bm{\sigma}$.
    \item {\tt DiracToWeyl[exp]} converts Dirac spinors and gamma matrices in {\tt exp} to Weyl spinors and Pauli matrices respectively.
    \item {\tt ContractNR} is the non-relativistic version of {\tt ContractLI}.
    \item {\tt SigmaReduce} reduces chains of Pauli matrices.
\end{itemize}

\subsection{Interfaces}\label{subsec:Int}

Interfaces to packages {\tt QGRAF}~\cite{Nogueira:1991ex}, FORM~\cite{Vermaseren:2000nd,Ruijl:2017dtg}, and {\tt Kira}~\cite{Maierhofer:2017gsa,Klappert:2020nbg} are provided. The usages are as follows.

\begin{itemize}
    \item {\tt ARQgraf[i->o, n]} generates the {\tt n}-loop amplitudes (or diagrams) for the process {\tt i->o}, with {\tt i} and {\tt o} the incoming and outgoing particles respectively. To use {\tt ARQgraf}, a Fortran compiler (like {\tt gfortran}) should be available. {\tt ARQgraf} uses {\tt parallel}~\cite{tange2018gnu} to parallelize computations (similarly for the other interfaces).\\
    For example, we can generate the one-loop amplitude for Bhabha scattering by running:\\
    \mathina~{\tt ARQgraf[\{"F", "F\_m"\} -> \{"F", "F\_m"\}, 1, "Model" -> "QED"]}\\
    {\tt ARQgraf} has the following options:\\
    {\tt "Style"} specifies whether the diagrams or the amplitudes are generated. The option value should be either {\tt "Diagram"} or {\tt "Amplitude"}. If {\tt "Style"->"Diagram"}, the generated diagrams can be painted by the function {\tt ARTikzFeynman}, which uses {\tt lualatex} together with {\tt tikz-feynman}~\cite{Ellis:2016jkw} to paint Feynman diagrams.\\
    {\tt "Model"} specifies the model file of {\tt QGRAF}. The built-in models are {\tt "QED"}, {\tt "QCD"}, and {\tt "WL"} (Wilson lines). In all these models, fermions are represented by {\tt "F\_"}. In the model {\tt "QED"}, photons are represented by {\tt "V\_"}. In the model {\tt "QCD"}, gluons are represented by {\tt "V\_1\_"} and photons are represented by {\tt "V\_2\_"}. The model {\tt "QCD"} is included in the model {\tt "WL"}. In the model {\tt "WL"}, a Wilson line is represented by {\tt "W\_"}. The anti-particle of a particle, say, a fermion, is represented by {\tt "F\_m"}. User-defined model files are also allowed. For user-defined model files, the option value of {\tt "Model"} should either be the file name or the file context as a string. The file name should be wrapped with single quotes, for example, {\tt "'./model.mod'"}. Two model files can be combined by putting them in a single list. For example, {\tt "Model"->\{"QCD","WL"\}}.\\
    {\tt "FeynmanRules"} specifies the Feynman rules. Similar to the option {\tt "Model"},\\
    {\tt "FeynmanRules"} could be either a file name or the file context. The built-in Feynman rules are {\tt "QED"}, {\tt "QCD"}, {\tt "QCDRXi"} (QCD in the general $R_{\xi}$ gauge), and {\tt "WL"}. User-defined Feynman rules are also allowed. {\tt "FeynmanRules"} is in fact some FORM codes. If some symbols (indices, vectors, functions) are present in {\tt "FeynmanRules"}, they should be specified through the option {\tt "Symbols"} ({\tt "Indices"}, {\tt "Vectors"}, {\tt "CFunctions"}).\\
    {\tt "ImportResults"} specifies whether to import the final results or not. If {\tt "ImportResults"\\
    ->False}, the results will be written to some files, and the file names will be returned. For processes with a large number of diagrams, it is suggested to set {\tt "ImportResults"\\
    ->False} to save memory.\\
    {\tt Prepend} is a list of FORM codes to prepend to the Feynman rules. And {\tt Append} is a list of FORM codes to append to the Feynman rules.\\
    {\tt Select} is a list of integers to select the specified diagrams.
    \item {\tt DoForm} is an interface to FORM, which uses FORM to calculate traces of gamma matrices and to contract Lorentz indices. To use {\tt DoForm}, one should make sure that FORM is correctly installed. {\tt DoForm} has the following options:\\
    {\tt Collect} is a list of patterns. If the option value is nonempty, terms not free of these patterns will be wrapped with the head {\tt HoldAR}.\\
    {\tt "Replacements"} is a list of rules for replacements that will be carried out during the calculation.\\
    {\tt "AnticommutativeGamma5"} specifies whether to use the anticommutative $\gamma^5$ scheme or not. If {\tt "AnticommutativeGamma5"->False}, $\gamma^5$ will be expressed in terms of the Levi-Civita tensor.\\
    {\tt "Executable"} specifies the executable to FORM.\\
    {\tt Path} is a list of directories where the executable can be found.
    \item {\tt DoKira} is an interface to {\tt Kira}, which uses {\tt Kira} to carry out the IBP reduction. To use {\tt DoKira}, one needs to append {\tt "UseKira"->True} to {\tt \$AROptions} before loading \ar. Besides {\tt Kira}, one should ensure that {\bf GiNaC}~\cite{Bauer:2000cp} is available, since {\tt DoKira} uses it to prepare IBPs.\\
    {\tt DoKira} has the following options:\\
    {\tt "UserDefinedSystem"} specifies whether to use the user-defined system of {\tt Kira} or not. Note that if {\tt "UserDefinedSystem"->True}, {\tt DoKira} uses the user-defined system of {\tt Kira} to solve IBP identities by brute force, and the symbolic rules are NOT implemented.\\
    {\tt Method} specifies the method to generate IBP identities. This option works only when {\tt "UserDefinedSystem"->True}.\\
    {\tt Bounds} specifies the bounds of indices while generating IBP identities. The option value should be a list of two integers (for the positive and negative indices respectively).\\
    {\tt "Executable"} specifies the {\tt Kira} executable together with the options.\\
    {\tt Directory} is the directory for the {\tt Kira} job files.\\
    {\tt "Compiler"} specifies the C++ compiler. And {\tt "CompilerFlags"} is a list of flags for the compiler. By default, {\tt "Compiler"->"g++"} and {\tt "CompilerFlags"->\{"-lginac","-ldl"\}}.\\
    {\tt "KiraJobFileOptions"} is a list of options for the {\tt Kira} job file.\\
    {\tt ClearKira} specifies whether to clear the working directory before creating the job files or not.\\
    {\tt "PrepareKira"} specifies whether to prepare the {\tt Kira} job files or not.\\
    {\tt "RunKira"} specifies whether to run the {\tt Kira} executable or not.\\
    {\tt "ImportResults"} specifies whether to import the reduction result or not.\\
    {\tt "IBP"} is a list of IBP identities. This option allows users to add IBP identities manually.\\
    {\tt "preferred\_masters"} is a list of preferred master integrals.
    \item {\tt ClearKira[]} clears the temporary files created by {\tt DoKira}.
\end{itemize}

\subsection{Auxiliary functions}\label{subsec:AuxFunct}

Some auxiliary functions are provided in the file {\tt auxiliary.m}. All these functions are defined in the {\tt "Global`"} context. These functions are not essential to \ar. Thus users can freely modify this file. The auxiliary functions are:

\begin{itemize}
    \item {\tt D2eps} expresses the spacetime dimension in terms of {\tt eps}.
    \item {\tt epsSeries[exp]} expands {\tt exp} in power series of {\tt eps}.
    \item {\tt FA2AR[amp]} converts an amplitude {\tt amp} generated by FeynArts~\cite{Kublbeck:1990xc,Hahn:2000kx} to the one of \ar style.
    \item {\tt SingularGroebnerbasis} is an interface to Singular~\cite{Greuel:2008}, which uses Singular to generate Gr\"obner bases.
    \item {\tt SingularPolynomialReduce} uses Singular to reduce polynomials.
    \item {\tt MatchBoundaryConditions[asy, bnd, x, C]}, with {\tt asy} the asymptotic solutions of a differential-equation system, {\tt bnd} the boundary conditions~(of the format \{\{i,o\}->c,\dots\}), and {\tt C} a pattern, determines the values of constants in {\tt asy} matching the pattern {\tt C} by matching the asymptotic solutions {\tt asy} to the boundary conditions {\tt bnd}.
\end{itemize}

\subsection{Examples}\label{subsec:Exampl}

Some examples on using \ar are provided in the folder {\tt AmpRed/examples}:
\begin{itemize}
    \item For beginners, it is suggested to start with the examples {\tt "Box.nb"}, \\
    {\tt "GeneralParametricIntegral.nb"}, and {\tt "HiggsDecay.nb"}.
    \item For the reduction of integrals, see the example {\tt "DoubleBox.nb"}.
    \item For the numerical calculation of master integrals, see the example {\tt "MasterIntegral.nb"}.
    \item For the application of the differential-equation method, see the example {\tt "Banana.nb"}.
    \item For the reduction of phase-space integrals, see the examples {\tt "CutIntegral.nb"} and {\tt "FourLeptonDecay.nb"}.
    \item For the combination of \ar with the other packages, see the examples \\
    {\tt "HiggsDecay\_TwoLoop\_QGRAF+FOMR+KIRA.nb"}, {\tt "EtacDecay.nb"}, and {\tt "EtacDecay\_2.nb"}.
\end{itemize}
Some of these examples are described in detail in this section.

\subsubsection{Reduction and numerical evaluation}
In this subsection, we introduce functions to reduce integrals and numerically evaluate master integrals through a simple double-box example. We do the reduction by using both the built-in function {\tt AlphaReduce} and the {\tt Kira} interface.

The package can be loaded by running
\begin{align*}
    \mathinb~&\texttt{AROptions=\{"UserKira"->True\};}\\
    &\texttt{<<AmpRed`}
\end{align*}
Before the calculations, we should set the kinematics first. That is, we express Lorentz invariants in terms of independent Mandelstam variables. And to parallelize the calculations, we launch some kernels.
\begin{align*}
    \mathinb~&\texttt{SetKinematics[\{p[1], p[2]\} -> \{k[1], k[2]\}, \{0, 0\} -> \{0, 0\}];}\\
    &\texttt{LaunchKernels[];}
\end{align*}
Here the first argument of {\tt SetKinematics} specifies the incoming and outgoing momenta, and the second argument specifies the corresponding masses. Here we set all the masses to $0$. {\tt SetKinematics} automatically sets Lorentz invariants in terms of Mandelstam variables $s$ and $t$. The symbols for Mandelstam variables can be changed by setting the option {\tt "Mandelstam"}. Alternatively, one can set the kinematics manually~(See the introduction to {\tt Pair} and {\tt SP} in sec.~\ref{sec:TensAlg}.)

We define a standard tensor integral by using {\tt TI}~(See sec.~\ref{subsec:FeynmInt}).
\begin{align*}
    \mathinb~&\texttt{int = TI[\{l[1], l[2]\}, \{\{k[1] + l[1]\}, \{k[1] + l[1] + l[2]\},}\\
    &\qquad\texttt{\{k[2] - l[1]\}, \{k[2] - l[1] - l[2]\}, \{k[2] - l[1] - l[2] - p[1]\},}\\
    &\qquad\texttt{\{l[1]\}, \{l[2]\}\}, FV[l[1], a]*FV[l[1], b]]}\\
    \mathoutb{~}&-\frac{\pi ^{-D}}{\Gamma (\epsilon +1)^2}\int d^Dl_1d^Dl_2\\
    &\frac{l_1^a l_1^b}{l_1^2 l_2^2 \left(k_2-l_1\right)^2 \left(k_1+l_1\right)^2 \left(k_2-l_1-l_2\right)^2 \left(k_1+l_1+l_2\right)^2 \left(k_2-l_1-l_2-p_1\right)^2}
\end{align*}

The integral reduction can be carried out by using the built-in function {\tt AlphaReduce}.
\begin{align*}
    \mathinb~\texttt{res1=AlphaReduce[int];}
\end{align*}
{\tt AlphaReduce} implements both method I and Method II described in sec.~\ref{sec:IntRed}. The method can be specified through the option {\tt Method}. By default, method II is used. Alternatively, the integrals can be reduced by using the {\tt Kira} interface {\tt DoKira}.
\begin{align*}
    \mathinb~\texttt{res2=DoKira[int];}
\end{align*}
Internally, {\tt DoKira} uses {\tt AlphaParametrize} to parameterize tensor integrals in terms of parametric integrals {\tt AlphaInt[...]} and then reduces parametric integrals by using {\tt Kira}. By default, {\tt DoKira} chooses the reduction method automatically according to the structure of the expression. Specifically, if all integrals in the expression are standard loop integrals, {\tt DoKira} uses the built-in reduction system of {\tt Kira} to carry out the reduction. If the integrals are with momentum-space correspondences but non-standard (that is, integrals with delta and Heaviside theta functions), the reduction will be carried out by using the user-defined system of {\tt Kira}, with the IBP identities generated by eq.~(\ref{eq:IBP3}). For integrals with no momentum-space correspondences, the reduction is carried out by the user-defined system with IBPs generated by eq.~(\ref{eq:IBP2}). One can also choose the method manually by setting the options {\tt "UserDefinedSystem"} and {\tt Method}.

The master integrals are numerically evaluated by using the function {\tt AlphaIntEvaluateN}. In principle, one can directly apply {\tt AlphaIntEvaluateN} to the original expression {\tt int}. Nevertheless, in practical calculations, for a better performance, it is suggested to carry out the calculations step by step. Here, since {\tt int} is already reduced to {\tt res1}, we can evaluate {\tt res1} instead.
\begin{align*}
    \mathinb~\texttt{resn=AlphaIntEvaluateN[res1,8,\{t->-1/3\}]};
\end{align*}
{\tt AlphaIntEvaluateN} returns the numerical result of the expression, together with a list of numerical results of all master integrals. By default, {\tt AlphaIntEvaluateN} uses {\tt AlphaReduce} to carry out the integral reduction, which can be changed by setting the option {\tt AlphaReduce}. For example, one can use {\tt DoKira} to do the integral reduction by adding an option\\
{\tt AlphaReduce->DoKira}.

\subsubsection{Differential equations}

In this subsection, we use a simple massive-banana example to show how to implement the differential equation method with \ar.

Again, we set the kinematics first.
\begin{align*}
\mathinb~&\texttt{SP[p, p] = x;}\\
&\texttt{m = 1;}
\end{align*}
We use {\tt DoKira} to do the integral reduction.
\begin{align*}
\mathinb~&\texttt{SetOptions[DoKira, \{"UserDefinedSystem" -> True, Method -> 2\}];}\\
&\texttt{SetOptions[AlphaDES, \{Method -> 2, AlphaReduce -> DoKira\}];}
\end{align*}
The integral to be calculated is
\begin{align*}
\mathinb~&\texttt{int = TI[\{l[1], l[2],  l[3]\}, \{\{l[1], m\}, \{l[2], m\}, \{l[3], m\},}\\
&\qquad\texttt{\{l[1] + l[2] + l[3] + p, m\}\}]}\\
\mathoutb~&\frac{i \pi ^{-3 D/2}}{\Gamma (\epsilon +1)^3}\int d^Dl_1d^Dl_2d^Dl_3\frac{1}{\left(l_1{}^2-1\right) \left(l_2{}^2-1\right) \left(l_3{}^2-1\right) \left(\left(l_1+l_2+l_3+p\right){}^2-1\right)}
\end{align*}
We use {\tt AlphaDES} to construct the differential equation system.
\begin{align*}
\mathinb~\texttt{des = AlphaDES[I*int, x];}
\end{align*}
We choose the boundary of the differential equations to be at $x=0$, which is in the Euclidean region. The boundary conditions of the differential-equation system can be determined by matching the asymptotic solutions of the differential-equation system with the asymptotic expansions of the master integrals. The former can be obtained by using {\tt DESAsymptoticSolve}, and the latter can be obtained by using {\tt AlphaSeries}. These operations are implemented in the function {\tt DESBoundary}.
\begin{align*}
\mathinb~\texttt{bnd = DESBoundary[des, x];}
\end{align*}
The boundary conditions are expressed in terms of parametric integrals, which can be numerically evaluated using {\tt AlphaIntEvaluateN}.
\begin{align*}
\mathinb~&\texttt{bndn = AlphaIntEvaluateN[Last /@ bnd, 12, \{\}, PrecisionGoal -> 40,}\\
&\qquad \texttt{AlphaReduce -> DoKira];}
\end{align*}
With the boundary conditions fixed, we can solve the differential equations numerically using the function {\tt DESSolvN}. Notice that {\tt DESSolveN} allows us to solve the differential equations at an arbitrary point, including the singularities. For example, we may solve the differential equations at the singularity $x=4$ by running
\begin{align*}
\mathinb~&\texttt{sol = DESSolveN[des[[2]], \{x, 0, 4\}, 8, 
    MapThread[Rule, \{First /@ bnd, }\\
    &\qquad\texttt{bndn[[1]]*eps\^~4\}], 
    PrecisionGoal -> 20]/eps\^~4;}
\end{align*}
Here the boundary constants are multiplied by a factor of {\tt eps\^~4} to make them free of poles in {\tt eps}.

\subsubsection{Calculation of a full amplitude}
In this subsection, we demonstrate how to use \ar to calculate a full amplitude by calculating the two-loop correction to Higgs two-photon decay.

For simplicity, we set all Lorentz invariants to numbers.
\begin{align*}
\mathinb~&\texttt{LaunchKernels[];}\\
&\texttt{SetKinematics[\{P\} -> \{p[1], p[2]\}, \{FA["MH"]\} -> \{0, 0\}];}\\
&\texttt{FA["MH"] = FA["EL"] = 1;}\\
&\texttt{FA["MT"] = 2;}\\
&\texttt{FA["MW"] = 1/2;}\\
&\texttt{FA["SW"] = 1/2;}
\end{align*}
We generate the amplitude using {\bf FeynArts}, which is saved in the file\\
{\tt "AmpRed/examples/HiggsDecay\_TwoLoop\_Amplitude.m"}. We can directly import it and convert it to the format of \ar using the function {\tt FA2AR}.
\begin{align*}
\mathinb~&\texttt{amp=FA2AR[Get["HiggsDecay\_TwoLoop\_Amplitude.m"],ToFeynmanInt->True]/.}\\
&\texttt{\{\_Incoming->P,Outgoing->p\}//SpinorChainSimplify//ColourSimplify;}
\end{align*}
Here the function {\tt SpinorChainSimplify} calculates all the traces of Dirac gamma matrices, and {\tt ColourSimplify} simplifies chains of color matrices.
We use the built-in function {\tt AlphaReduce} to reduce the integrals.
\begin{align*}
\mathinb~\texttt{amp1 = AlphaReduce[amp];}
\end{align*}
We can check that the result satisfies the Ward identity.
\begin{align*}
\mathinb~&\texttt{Total[amp1]/.PolarizationVector[p[1],\_\_]->p[1]//ContractLI//Simplify}\\
  \mathoutb~&0
\end{align*}
Finally we calculate the master integrals numerically by
\begin{align*}
\mathinb~\texttt{amp2 = AlphaIntEvaluateN[amp1/.ColourN["F"]->3, 8, \{\}, PrecisionGoal->20];}
\end{align*}

\section{Summary and discussion}

In this paper, we present \ar, a Mathematica package for the semi-automatic calculations of multi-loop Feynman amplitudes. \ar implements the methods developed in refs.~\cite{Chen:2019mqc,Chen:2019fzm,Chen:2020wsh} to reduce Feynman integrals through the parametric representation and the method developed in ref.~\cite{Chen:2023hmk} to calculate parametric integrals iteratively. These methods are briefly reviewed in this paper. The detailed usages of \ar are introduced. \ar allows for full calculations of amplitudes: tensor algebras, reduction of amplitudes, and the numerical calculations of master integrals. Various user-friendly tools for multi-loop calculations are provided. It also provides interfaces to packages including {\tt QGRAF}, FORM, and {\tt Kira}.

\ar uses the algorithms (method I and method II) described in ref.~\cite{Chen:2019fzm} to reduce tensor integrals. In method I, tensor integrals are directly parametrized and are expressed in terms of parametric integrals with negative indices. The IBP identities for the resulting parametric integrals are equivalent to the traditional momentum-space IBP identities. In method II, tensor integrals are first parametrized and partially reduced by using the techniques of polynomial reduction based on eq.~(\ref{eq:IBPB}). Then the unreduced integrals are further reduced by combining symbolic rules with a Laporta-like algorithm.

Since the IBP identities used in method I are equivalent to the traditional ones, it allows users to first parametrize tensor integrals by using \ar and carry out the IBP reduction by using some user-defined reducers.

Reduction with method II is much more efficient than that with method I. Currently, this method is only implemented in a pure Mathematica code. While implementing the full method with another language other than Mathematica is far from trivial, parts of it are expected to be implemented in the future, such as the symbolic rules. An interface to {\tt Kira} is provided, but \ar only uses the user-defined system of {\tt Kira} to solve IBP identities generated by eq.~(\ref{eq:IBP2}) by brute force, and the symbolic rules are not implemented. An observation is that even if we solve IBP identities by brute force, the efficiency is comparable with traditional methods while generating differential-equation systems.

\ar uses the method developed in ref.~\cite{Chen:2023hmk} to calculate master integrals iteratively. For real integrals, this method proves to be more efficient than other methods on the market in most circumstances. However, for complex integrals, it sometimes becomes time-consuming to use this method. Thus, for complex integrals, it is suggested to calculate them by solving differential equations (by using the function {\tt DESSolveN}) and calculate the boundary conditions (chosen in the Euclidean region) by using this method (with the function {\tt AlphaIntEvaluateN}). It is expected to refine the method of analytic continuation in the future.

Currently, the iterative method does not work for phase-space integrals. The main obstacle is the lack of a systematic algorithm for the asymptotic expansions of phase-space integrals. It is also expected to solve this problem in the future.

\acknowledgments

This work is supported by National Natural Science Foundation of China~(NNSFC) under Grant No. 12405095 and Guangdong Major Project of Basic and Applied Basic Research~(No. 2020B0301030008).





\bibliographystyle{JHEP}
\bibliography{biblio.bib}

\end{document}